\documentclass[journal]{IEEEtran}

\usepackage{cite}

\usepackage{amsmath,amsthm}
\DeclareMathOperator*{\argmin}{arg\,min}
\DeclareMathOperator*{\argmax}{arg\,max}

\newcommand{\myedit}[1]{{\color{black}#1}}

\usepackage{mathrsfs}
\usepackage{hyperref}
\usepackage{algorithm}
\usepackage{algpseudocode}
\makeatletter
\def\BState{\State\hskip-\ALG@thistlm}
\makeatother

\algdef{SE}[SUBALG]{Indent}{EndIndent}{}{\algorithmicend\ }%
\algtext*{Indent}
\algtext*{EndIndent}

\usepackage{amssymb}
\usepackage{graphicx}
\usepackage{float}
\usepackage{color}

\begin{document}

\thispagestyle{empty} 
\noindent
\begin{center}
    \large\textbf{Copyright Notice}
\end{center}

\vspace{2cm}

\noindent
\copyright\ 2019 IEEE. Personal use of this material is permitted. Permission from IEEE must be obtained for all other uses, in any current or future media, including reprinting/republishing this material for advertising or promotional purposes, creating new collective works, for resale or redistribution to servers or lists, or reuse of any copyrighted component of this work in other works.

\vspace{1cm}

\noindent
\textbf{To cite this article:} \\
S. Lin, ``Robust Pitch Estimation and Tracking for Speakers Based on Subband Encoding and the Generalized Labeled Multi-Bernoulli Filter,'' in \textit{IEEE/ACM Transactions on Audio, Speech, and Language Processing}, vol. 27, no. 4, pp. 827-841, 2019.

\vspace{1cm}

\noindent
\textbf{Official Version of Record:} \\
The final version of this article is available at: 

\url{https://ieeexplore.ieee.org/abstract/document/8638997}

\url{https://doi.org/10.1109/TASLP.2019.2898818}

\vspace{2cm}
\noindent
\textbf{Dr. Shoufeng Lin has been a Senior Member of IEEE since 2020.}

\newpage 
\setcounter{page}{1} 

\title{Robust Pitch Estimation and Tracking for Speakers Based on Subband Encoding and the Generalized Labeled Multi-Bernoulli Filter}

\author{Shoufeng~Lin,~\IEEEmembership{Member,~IEEE}
\thanks{Shoufeng Lin is currently
with the School of Electrical Engineering, Computing and Mathematical Sciences, Curtin University, Bentley, Western Australia. E-mail: shoufeng.lin@postgrad.curtin.edu.au; ee.linsf@gmail.com.}
}

\maketitle

\begin{abstract}
This paper proposes a new pitch estimator and a novel pitch tracker for speakers. We first decompose the sound signal into subbands using an auditory filterbank, assuming time-frequency sparsity of human speech. 
Instead of directly selecting the number of subbands according to experience, we propose a novel frequency coverage metric to derive the number of subbands and the center frequencies of the filterbank. 
The subband signals are then encoded inspired by the computational auditory scene analysis (CASA) approach, and the normalized autocorrelations are calculated for pitch estimation. To suppress spurious errors and track the speaker identity, the temporal continuity constraint is exploited and a Generalized Labeled Multi-Bernoulli (GLMB) filter is adapted for pitch tracking, where we use a novel pitch state transition model based on the Ornstein-Uhlenbeck process, and the measurement driven birth model for adaptive new births of pitch targets. 
Experimental evaluations with various additive noises demonstrate that the proposed methods have achieved better accuracy compared with several state-of-the-art pitch estimation methods in most studied scenarios. Tests using real recordings in a reverberant room also show that the proposed method \myedit{is} robust against reverberation.
\end{abstract}


\IEEEpeerreviewmaketitle

\section{Introduction}

\IEEEPARstart{P}{itch} estimation and tracking can play an important part in many audio signal processing applications including automatic speaker identification, speech separation and transcription. 
In this paper, we focus on extracting fundamental frequencies\footnote{In this paper, we use ``pitch'' and ``fundamental frequency'' interchangeably.} of human speakers, from single channel sound signals, which can be speech signals from a single speaker or concurrent speakers, mixed with noises or reverberation from the environment.

Many efforts have been made in estimating the fundamental frequency of voiced sound signals. 
Time domain methods investigate the periodic patterns of signals, and often apply the autocorrelation function (ACF), cross-correlation function (CCF), average magnitude difference function (AMDF) or the cumulative mean normalized difference function, etc. to the sound signals to detect the time delays that correspond to the fundamental periods {\color{black}\cite{ross1974average, talkin1995robust, meddis1997unitary, de2002yin, garner2013simple}}. 
Frequency domain methods study the harmonic structure of sound signal spectra and extract pitch information based on various features and rules, e.g. the harmonic product spectrum \cite{noll1970pitch}, subharmonic summation \cite{hermes1988measurement}, {\color{black}wavelet based instantaneous frequency \cite{kawahara1999fixed, kawahara1999restructuring}} and the subharmonic-to-harmonic ratio \cite{sun2002pitch}, etc. 
Most of the existing methods can produce reliable pitch estimation results in amiable environments, but strong noises or reverberation can degrade the performance significantly, by corrupting the periodic patterns of time domain signals or the harmonic structures of the signal spectra. 
Other recent advances on robust pitch estimation include mainly the more complicated features and strategies, e.g. total energy of harmonics \cite{gonzalez2014pefac}, harmonic frequency deviation \cite{wang2017robust}, etc. 
The subspace-based method \cite{christensen2007joint, christensen2008robust, zhang2010robust} have been developed to decouple speech and noise subspaces and can provide high resolution pitch estimates. 
Some more statistical methods provide probabilistic models for noisy sound signals and find pitch estimates with optimal probabilities according to their models and the observations \cite{tabrikian2004maximum, chu2012safe}. 
However, a majority of existing methods are designed for the pitch estimation of a single speaker.

For co-channel multi-pitch estimation of concurrent speakers, several works inspired by the computational auditory scene analysis (CASA) approaches (e.g. \cite{CASAwang}) have been developed \cite{tolonen2000computationally, wu2003multipitch}. They work on the time-frequency (TF) domain by decomposing single-channel sound signal into subbands via an auditory filterbank and then performing time-domain analysis in each subband. 
Although the center frequencies of subband filters of an auditory filterbank can be derived by selecting a number of subbands equidistantly on a frequency scale, e.g. the logarithmic scale \cite{sun2002pitch}, Bark scale \cite{smith1999bark} or the ERB-rate scale\cite{glasberg1990derivation,rouat1997pitch, wu2003multipitch}, etc., the selection of the total number of subbands has been essentially empirical. Apparently, using more subbands than necessary can impair computational efficiency, while insufficient subbands can cause loss of information and hence estimation errors. 
%
Furthermore, to obtain continuous pitch contours, temporal continuity constraint of pitch is often exploited, assuming continuous speech production by the human vocal system. 
Several pitch tracking methods based on the hidden Markov model (HMM) \cite{wu2003multipitch, lee2012noise, wang2017robust},{\color{black} \cite{reddy2017robust}} can be found in the literature, forming pitch tracks via estimating the hidden state sequence from observations. 
A recent work \cite{wohlmayr2011probabilistic} uses trained Gaussian mixture models (GMMs) for signal spectrogram and put the probabilistic speaker observation models under the factorial hidden Markov models (FHMM) framework for multi-pitch tracking. 
{\color{black}Neural networks (NN) form another emerging approach for pitch estimation and tracking, using various features, neurons and network topologies \cite{wang2014f0, zhang2016rnn, liu2017speaker}.}
However, for accurate pitch {\color{black} estimation and} tracking results, the HMM {\color{black}and NN} based methods usually require carefully training the algorithms to obtain accurate hidden state transition probabilities, which can be inconvenient and restricting in practice.


In this paper, we first propose a novel speaker pitch estimator. It uses an auditory filterbank to decompose speech signals into subbands, based on the TF sparsity assumption \cite{yilmaz2004blind} of speech signals. The number of subbands (and hence center frequencies) of the filterbank is calculated according to our proposed ``frequency coverage'' metric for consistent and full coverage over the frequency range. 
Moreover, inspired by the CASA approach and psychoacoustic studies, we propose to encode the subband signals with a robust encoding model to obtain distinct and reliable pitch estimates for the possibly noisy and quasi-periodic speech signals. 
Pitch estimates are then selected from the normalized autocorrelation coefficients of the encoded subband signals. {\color{black}Some preliminary results can be found in \cite{lin2018new}, and this paper provides detailed derivations and further evaluations.}

We also propose a novel ``training-free'' pitch tracker based on the Generalized Labeled Multi-Bernoulli (GLMB) filter \cite{vo2013labeled, reuter2014labeled, vo2014labeled}, to further reduce spurious estimation errors and to track pitch estimates with identities (i.e. to associate the pitch estimate with the corresponding speaker). The GLMB filter has been successful in tracking locations of multiple targets, but necessary adaptations are required for the pitch tracking problem. 
We propose to model the pitch state transition as an Ornstein Uhlenbeck process \cite{gardiner1985handbook}, assuming temporal continuity of speech production and that the pitch of a speaker tends to return to its average level. We also apply the measurement driven birth model for the adaptive new births of pitch targets in the GLMB prediction steps \cite{reuter2014labeled, lin2016measurement}, and provide the adaptations for the cases of a single pitch target and long pauses during speech. 
The resulting pitch tracking filter also assigns a unique identity to pitch estimates of a corresponding speaker, and can thus form linked tracks of pitch estimates for the respective speakers over time. This novel pitch tracker is applicable to tracking pitches of a single speaker, as well as concurrent speakers with pitches at different levels. It uses basic generic statistic models for the pitch state transition and observations, and therefore does not require training.

The rest of the paper is organized as follows. Section \ref{sec:pitch-estimation} presents our novel pitch estimator, which is followed by the proposed pitch tracker in Section \ref{sec:pitch-tracking}. Numerical studies are carried out in Section \ref{sec:performance}, and conclusions are provided in Section \ref{sec:conclusion}.

\section{Speaker Pitch Estimator}
\label{sec:pitch-estimation}

\subsection{Speech Signal Model}
In a noisy and reverberant environment, sound signal acquired by a single microphone is a mixture of reverberated speech signal from the speaker(s) and noise: 
\begin{equation} \label{eq:micSig}
{x}(t) 
= \sum \limits _{q=1} ^{Q} s_q(t) \ast {\mathrm{h}}_{q}(t)  + n(t)
,
\end{equation}
where $t\in \mathbb{R}$ is the continuous time,  the convolution operation is denoted as $\ast$, the additive noise at the microphone as $n(t)$. ${\mathrm{h}}_{q}(t) \in \mathbb{R}$ is the acoustic room impulse response (RIR) from the $q$-th speaker to the microphone. $q = 1,...,Q $, and integer $Q\geq 1$ the number of concurrent speakers. $s_q(t)$ is the sound signal from speaker $q$. {\color{black}The unvoiced part of $s_q(t)$ is often regarded as a stochastic process, while the voiced part can be} modeled based on the source excitation - vocal tract models for the process of speech production \cite{deller1993discrete}, and the amplitude-modulation (AM) and frequency modulation (FM) structure \cite{maragos1993energy},{\color{black}\cite{lin2018reverberation}}:
\begin{equation}
\label{eq:speechModel}
s_q(t) = \sum \limits _{{\hbar}=1} ^{H_q} A^{(\hbar)}_{q}(t) \cdot \cos \big( 2 \pi \cdot  {\hbar} \cdot f_q \cdot t + \phi^{({\hbar})}_{q}(t) \big) ,
\end{equation}
where integer ${\hbar}$ the order of harmonics for a speaker, integer $H_q$ the maximum order of harmonics for speaker $q$, $A^{({\hbar})}_{q}(t) \geq 0$ the envelope of each harmonic, $\phi^{({\hbar})}_{q}(t) \in \mathbb{R}$ the slow time-varying phase (which makes the speech signals quasi-periodic), $f_q > 0$ the desired fundamental frequency. Compared to the modulating harmonic frequency, the envelope $A^{({\hbar})}_{q}(t)$ is usually narrow band. Note that the amplitude of the fundamental frequency component may not be the strongest, due to the speech production process. 
%


\subsection{Subband Decomposition}
Based on the TF sparsity assumption \cite{yilmaz2004blind}, to separate harmonic components from the speaker(s), 
the microphone signal can be decomposed via an auditory filterbank{\color{black}\cite{CASAwang,lin2018reverberation}}:
\begin{equation}
\label{eq:micSignal}
{x}^{(b)}(t) = x(t) \ast g^{(b)}(t) ,
\end{equation}
where ${x}^{(b)}(t)$ denotes the decomposed signals from the microphone in subband $b$, $b=1,\dots, N_b$, integer $N_b$ is the total number of subbands, and 
$g^{(b)}(t)$ is the filter impulse response of subband $b$, which is aligned in time between subbands. Common auditory filters include the gammatone filter \cite{patterson1987efficient, holdsworth1988implementing, CASAwang}, gammachirp filter, etc. as well as their variants.
In this paper, we use the gammatone filter {\color{black}in \cite{holdsworth1988implementing}}, which can be expressed as
\begin{equation} \label{eq:gammatoneenv}
g^{(b)}(t) 
= \tilde{g}^{(b)}(t) \cdot \cos(2\pi f_C^{(b)} t ) , 
\end{equation}
where 
\begin{equation}
\tilde{g}^{(b)}(t) = ({t+t_d})^{\vartheta-1} e^{-2\pi f_b^{(b)} (t+t_d)} , 
\end{equation}
integer $\vartheta$ is the order of filter ($\vartheta=4$ in this paper), $t_d$ is time delay for alignment between filter bands, $f_b^{(b)}$ scaling factor for the bandwidth \cite{patterson1987efficient, CASAwang}, and $f_C^{(b)}$ is the center frequency of filter band $b$. 

From (\ref{eq:micSig}), (\ref{eq:speechModel}),  (\ref{eq:micSignal}), when the harmonic component $\hbar$ of the $q$-th speaker falls within the passband of subband $b$, and the noise is small in the particular subband, using the commutativity and associativity properties of convolution, and the frequency selectivity of the filterbank, the decomposed signals in subband $b$ become:
\begin{equation}
\begin{aligned} \label{eq:fbsig}
{x}^{(b)}(t)  
& = \big[ \sum \limits _{q=1} ^{Q} s_q(t) \ast \mathrm{h}_{q}(t)  + n(t) \big] \ast g^{(b)}(t)  \\
& \approx  \sum \limits _{q=1} ^{Q} s_q(t) \ast \mathrm{h}_{q}(t)  \ast g^{(b)}(t)  \\
%
\end{aligned}
\end{equation}
%
%
Assuming that the reverberation is not too strong, the RIR can be simplified as 
\begin{equation} \label{eq:RIR}
\mathrm{h}_{q}(t) = \mathrm{h}_q(t_{d_q}) \cdot \delta(t-t_{d_q}) ,
\end{equation}
where $\mathrm{h}_q(t_{d_q}) > 0$ and $t_{\color{black}d_q} > 0$ are the direct path amplitude and time-delay of the RIR from speaker $q$ to the microphone. 

Hence from (\ref{eq:gammatoneenv}), (\ref{eq:fbsig}) and (\ref{eq:RIR}), when the harmonic frequency falls in the subband, we have (see Appendix~\ref{appen:dirsubsig})
\begin{equation} \label{eq:subbandSignal}
{x}^{(b)}(t)  \approx  \tilde{S}_{q}^{(b)}(t)  \cdot \cos ( \tilde{\phi}^{({b})}_{q}(t)  )  ,
\end{equation}
where
\begin{equation}
\tilde{S}_{q}^{(b)}(t) =  \frac{1}{2} \cdot \mathrm{h}_{q}(t_{d_{q}})  \cdot 
  A^{(\hbar)}_{q}(t-t_{d_{q}}) \ast \tilde{g}^{(b)}(t)  , 
\end{equation}
%
%
\begin{equation} \label{eq:phib}
\tilde{\phi}^{({b})}_{q}(t) =   2\pi \hbar f_q \cdot (t-t_{d_{q}})  + \phi^{({\hbar})}_{q}(t-t_{d_{q}}) .
\end{equation}
%

\subsection{Frequency Range, Scale and Coverage}
\label{sec: freqCov}

A critical part of the subband approach in speech processing is the selection of center frequencies $f_C^{(b)} $ for subband filters $g^{(b)}(t)$, according to the chosen frequency range $[f_{min}, f_{max}]$, where $f_{max} >f_{min} > 0$. This is usually addressed by choosing a frequency scale and the corresponding number of subbands $N_b$. 
Various frequency scales have been used in the pitch estimation literature, including the logarithmic \cite{sun2002pitch}, Bark \cite{smith1999bark} and ERB-rate scales \cite{glasberg1990derivation}. 
However, in the current literature, the number of subbands for a given frequency scale in the given frequency range largely varies from one implementation to another, with no clear reason other than as an empirical choice. 
In \cite{rouat1997pitch}, a total of 20 subbands are used for frequency range of $330$Hz to $3700$Hz, while \cite{wu2003multipitch} implements 128 gammatone filters between $80$Hz and $5000$Hz.

In this paper, we use the ERB-rate scale (ERBS) as developed in \cite{glasberg1990derivation}. 
Denote the general form of ERB as
\begin{equation} \label{eq:linearERB}
{\upsilon} (f) = D + E \cdot f , 
\end{equation}
where $D = 24.7$, and $E = 0.108$ as given in \cite{glasberg1990derivation}.

From (\ref{eq:linearERB}), the resulting ERBS becomes {\color{black}(see Appendix~\ref{sec:ERBS})}:
\begin{equation} \label{eq:ERBS}
{\Upsilon}(f) \triangleq \int \frac{1}{{\upsilon}(f)} df 
 = E' \lg ( 1+ D' \cdot f ) ,
\end{equation}
with the boundary condition
${\Upsilon}(0) = 0 $. Here $D' \triangleq \frac{E}{D} $
and
$ E' \triangleq \frac{1}{E \cdot \lg e } $.
As given in \cite{glasberg1990derivation}, $E'= 21.4$, and $D'=0.00437$.

To derive the total number of subbands and the subband center frequencies for a given frequency range, we propose to use the ``frequency coverage'' metric, i.e. 
\begin{equation} \label{eq:etaCb}
\begin{aligned}
\eta_C^{(b)} 
& \triangleq \frac{ \frac{1}{2} \cdot (f_B^{(b+1)} + f_B^{(b)}) }{ f_C^{(b+1)} - f_C^{(b)} } , \\
\end{aligned}
\end{equation}
%
where $f_B^{(b)}$ denotes the filter bandwidth of subband $b$. 
\begin{figure}[!h]
\centering
\includegraphics[width=0.49\textwidth]{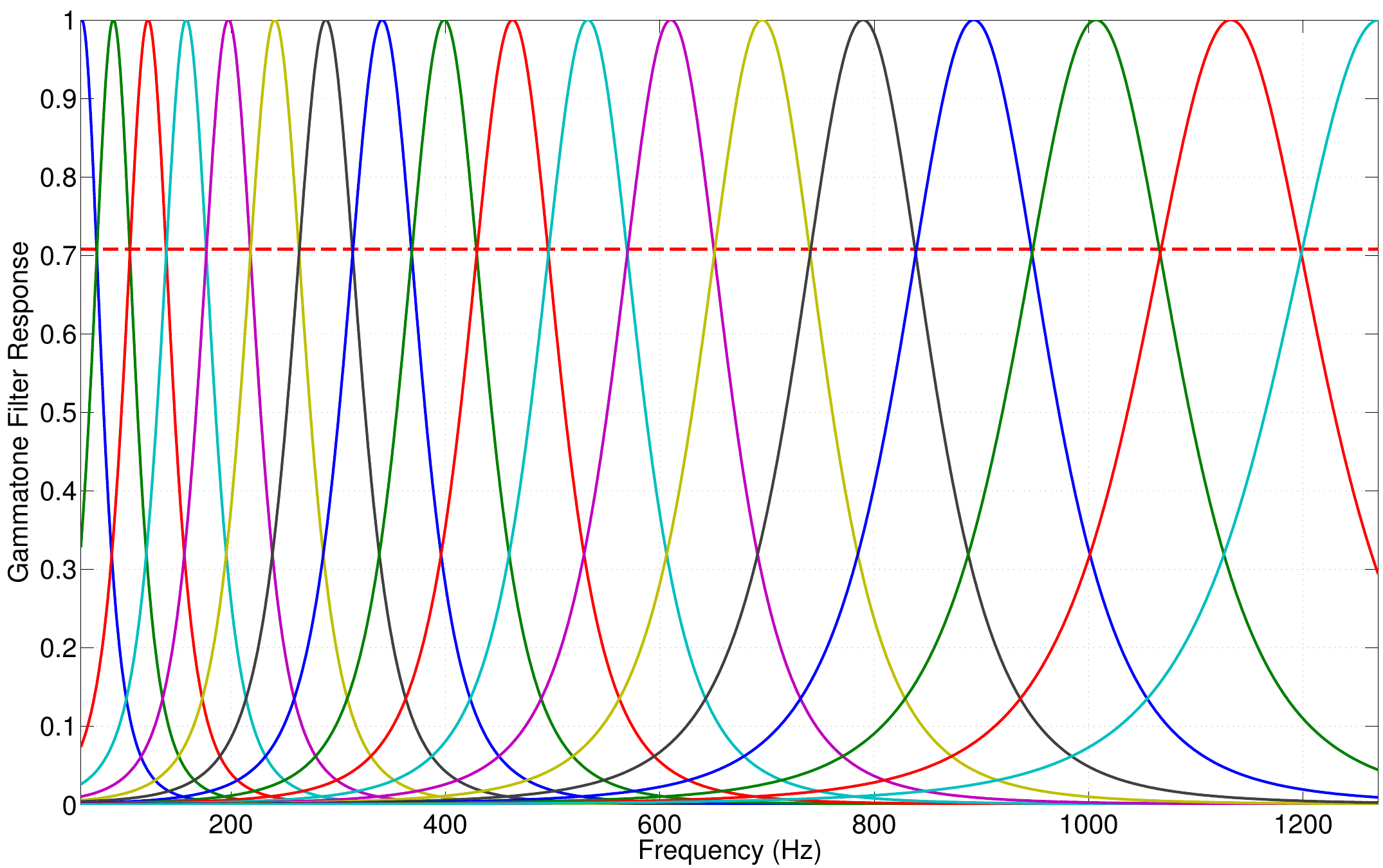}
\caption{An example of the frequency coverage metric (using the Gammatone filters). The frequency range is 60Hz to 1270Hz, thus for $\eta_c=1$ there are 18 subbands and the -3dB \myedit{passbands} of resulting subband filters align.} 
\label{fig:freqCov}
\end{figure}
{
As its name indicates, the ``frequency coverage'' metric measures how much of the frequency range is `covered' by all the \myedit{passbands} of subband filters.
This is easy to understand by first considering an ideal ``brick-wall'' bandpass filterbank.
Fig.~\ref{fig:freqCov} also shows the intuition for the proposed frequency coverage metric using the Gammatone filter. 
For $\eta_C=1$, the -3dB \myedit{passbands} of adjacent Gammatone filters align with no overlap. 
}
Apparently, a filterbank has consistent and full frequency coverage when $\eta_C^{(b)} \equiv 1$. For $\eta_C^{(b)} < 1$, there are some frequencies falling out of the passbands of the filterbank, which may result in estimation error when these frequencies include the desired fundamental frequency. 
The case of $\eta_C^{(b)}>1$ still leads to full frequency coverage, but there are redundancies as some frequency components are captured and analysed multiple times.
%
%

The linear relationship between bandwidth and center frequency holds for certain types of filters. Particularly, for the gammatone filter we have \cite{holdsworth1988implementing}:
\begin{equation} \label{eq:Kvartheta}
f_B^{(b)} = K_\vartheta \cdot f_b^{(b)} = K_\vartheta \cdot {\upsilon} (f_C^{(b)}) , 
\end{equation}
where $K_\vartheta$ is a constant for a given filter order $\vartheta$ as given in (\ref{eq:Ktheta}). In particular, $K_4=0.887$. {\color{black}Here $\vartheta=4$ is chosen for sufficient subband frequency selectivity (the attenuation is larger than 24dB at $f_C^{(b)} \pm \myedit{2} f_B^{(b)}$ for the 4th-order gammatone filter of subband $b$).}

The subband center frequencies in the given frequency range are distributed equidistantly on the ERBS, i.e.: 
\begin{equation} \label{eq:upsilon0}
f_C^{(b)} = \Upsilon^{-1} ( \frac{(N_b-b) \cdot \Upsilon(f_{min}) + (b-1) \cdot \Upsilon(f_{max})}{N_b-1} ) .
\end{equation}
%


Therefore the number of subbands $N_b$ can be derived from (\ref{eq:etaCb}), (\ref{eq:Kvartheta}) and (\ref{eq:upsilon0}) (see Appendix \ref{sec:appendix2}): 
\begin{equation} \label{eq:Nb}
N_b = \mathrm{round} \Big( 1 + \frac{ \ln (\frac{D + E\cdot f_{max}}{D + E \cdot f_{min}} ) } {\ln (\frac{{2\eta_C^{(b)}} + {E \cdot K_\vartheta}}{{2\eta_C^{(b)}} - {E \cdot K_\vartheta}} ) } \Big) . 
\end{equation}
This provides a consistent way for calculating the number of subbands in a given frequency range based on the frequency coverage metric. 
Once $N_b$ is obtained, the center frequencies can also be calculated from (\ref{eq:upsilon0}). In this paper, we choose $\eta_C^{(b)} \equiv 1$ for full frequency coverage without redundancies in processing. Since we keep $\eta_C^{(b)} $ the same for all subbands, $\eta_C$ is used hereafter for simplicity of denotation.

The pitch frequency range is denoted as $[F0_{min}, F0_{max}]$. In this paper, we choose $F0_{min} = 60$Hz, and $F0_{max}=500$Hz to cover the pitch range of most speakers \cite{deller1993discrete, nolan2003intonational}. 
Accordingly, the minimum subband frequency is chosen as $f_{min}= F0_{min} = 60$Hz. It has been pointed out that while low frequency auditory nerve fibers of inner hair cells tend to phase lock to pitch stimulus, those of frequencies above $1300$Hz do not \cite{rouat1997pitch}. Thus we choose $f_{max} = 1270$Hz in this paper \cite{rouat1997pitch}. 
Although autocorrelations of subband envelopes in frequencies higher than $1300$Hz were used in \cite{rouat1997pitch}, this high frequency range is not needed in our proposed method. 
Thus for $\eta_C=1$ we can get $N_b=18$ from (\ref{eq:Nb}) for the frequency range of $[60, 1270]$Hz.



\subsection{Rectification and Pitch Encoding}

In practice, signals are discretized at a sampling frequency of $f_s > 0$. We first half-wave rectify the discrete subband signal as in \cite{lyon1983computational,meddis1997unitary, tolonen2000computationally}:
\begin{equation}
\hat{x}^{(b)} (k/f_s) = \frac{1}{2} \cdot ({x}^{(b)} (k/f_s) + |{x}^{(b)} (k/f_s) | ) ,
\end{equation}
where the discrete time index $k \in \mathbb{Z}$.  

Assuming a slow-changing $\phi^{({\hbar})}_{q}(t)$ in (\ref{eq:subbandSignal}), we can rewrite the half-wave rectified subband signal as a convolution:
\begin{equation} \label{eq:pitchSignals}
\hat{x}^{(b)} (k/f_s) 
\approx 
\zeta_{\mathrm{cosine}}^{(\hbar,q)}(k)
\ast 
\sum_{\hat{k}_{n}^{(b)} \in \hat{K}^{(b)}}  \tilde{S}_{q}^{(b)}(k/f_s) \cdot \delta(k- \hat{k}_{n}^{(b)}), 
\end{equation}
where $\delta(\cdot)$ is the Dirac delta function, and $\zeta_{\mathrm{cosine}}^{(\hbar,q)}(k)$ is the non-negative part of the cosine term with peak at $k=0$, i.e. 
\begin{equation}
\begin{aligned}
\zeta_{\mathrm{cosine}}^{(\hbar,q)}(k) \triangleq &
%
\cos ( 2\pi \hbar f_q \cdot  k / f_s ) , k \in [- \frac{ f_s}{4  \hbar f_q} , \frac{ f_s}{4  \hbar f_q} ] , 
\end{aligned}
\end{equation}
$\hat{K}^{(b)} \triangleq \{ \hat{k}_{n}^{(b)} | ~n = 0,1,... \}$, and $\hat{k}_{n}^{(b)}$ is the index of a local peak
\begin{equation} \label{eq:localpeaks}
\hat{k}_{n}^{(b)} = \argmax\limits_{k}  \hat{x}^{(b)}(k/f_s) ,~ \forall~ k \in ({k}^{(b)}_{n-}, {k}^{(b)}_{n+}) ,
\end{equation}
${k}^{(b)}_{n-}, {k}^{(b)}_{n+}$ are consecutive zero-crossings of $\hat{x}^{(b)} (k/f_s)$ that satisfy
\begin{equation}
\hat{x}^{(b)} (k/f_s) > 0,~\forall~ k \in  ({k}^{(b)}_{n-}, {k}^{(b)}_{n+}) .
\end{equation} 

The speaker pitch can be found from the periodicity information of scaled delta functions, by searching for the peak of autocorrelation results, but the slow-changing cosine term can make the peak widespread or even cause spurious estimates. 
Actually we can check the time intervals between peaks of the scaled delta functions, i.e. $\tilde{S}_{q}^{(b)}(k/f_s) \cdot \delta(k- \hat{k}_{n}^{(b)})$ as in (\ref{eq:pitchSignals}). 
The problem is that the voiced speech signal is quasi-periodic, and the scaled delta functions alone can be sensitive to noise (the noise can affect the time indices of peaks), in the autocorrelation. 
Therefore, inspired by the approaches of computational auditory scene analysis (CASA) \cite{CASAwang, meddis1986simulation}, we propose to encode the subband signals as convolution of the scaled delta functions with a symmetrical encoding template, which in effect replaces the cosine term in (\ref{eq:pitchSignals}):
%
\begin{equation} \label{eq:pitchEncoding}
\zeta_p^{(\hbar,q)}(k) \triangleq 
\begin{cases}
e^{- |k|},~ k \in [-5,5] \\
0, ~ \mathrm{otherwise} ,
\end{cases}
\end{equation}
where we {\color{black}empirically} choose a fixed decay rate, 
so that the spike decays to $5\%$ of its peak in about $0.2$ms at a sampling rate of $f_s=16000$Hz in this paper. 
This aligns with the psychoacoustic observation of the exponential decay of the synaptic cleft contents from the hair cell in the organ of Corti\cite{meddis1986simulation}. 
The encoding template is symmetrical to avoid bias of time delay estimation in the autocorrelation. 
\begin{figure}[!h]
\centering
\includegraphics[width=0.49\textwidth]{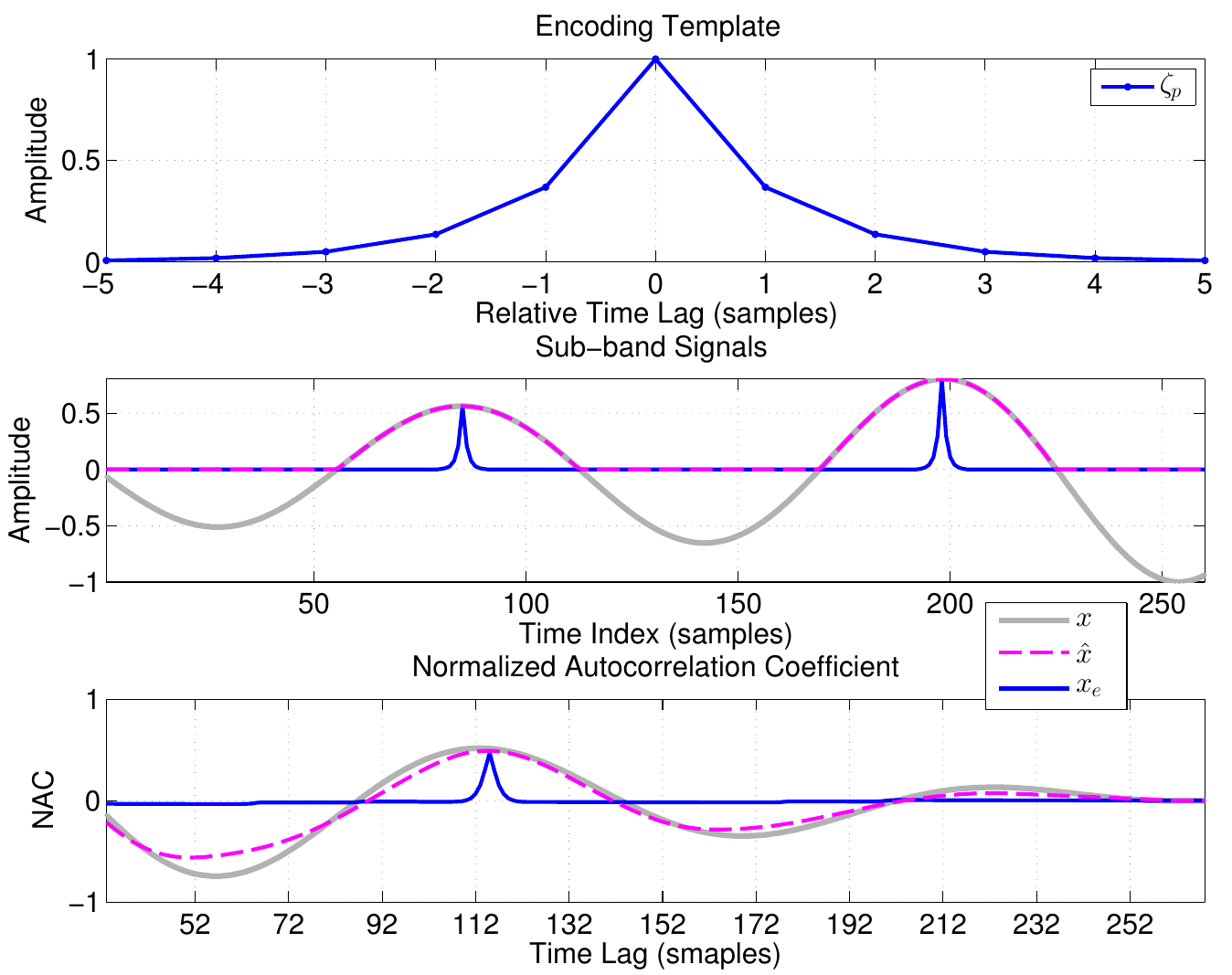}
\caption{Pitch encoding template (top panel), a subband signal from the filterbank, {\color{black}${x}^{(b)}$,}  its half-wave rectified {\color{black}$\hat{x}^{(b)}$}  and encoded signal (middle panel) {\color{black}${x}_e^{(b)}$}, and normalized autocorrelation coefficient of respective signals (bottom panel). }
\label{fig:pitch_encoded}
\end{figure}
Moreover, the encoding template can also be connected with the observation of the Laplacian distribution of peaks versus the relative time lags \cite{wu2003multipitch, chu2012safe}, except that for simplicity we discard (truncate) smaller values in (\ref{eq:pitchEncoding}) and assume that its dependence on subband indices {\color{black}and speakers} is negligible and the constant coefficient for the exponential term is $1$ as it does not affect the resulting normalized correlation coefficients. 

The resulting encoded subband signal from (\ref{eq:pitchSignals}) and (\ref{eq:pitchEncoding}) is
\begin{equation} \label{eq:encodedSubband}
x_e^{(b)}(k) = \zeta_p^{(\hbar,q)}(k) \ast \sum_{\hat{k}_{n}^{(b)} \in \hat{K}^{(b)}}  \tilde{S}_{q}^{(b)}(k/f_s) \cdot \delta(k- \hat{k}_{n}^{(b)}) .
\end{equation}
The top two panels of Fig.~\ref{fig:pitch_encoded} depict the encoding template, a segment of a subband signal {\color{black}${x}^{(b)}$} , its half-wave rectified signal {\color{black}$\hat{x}^{(b)}$} and its encoded signal {\color{black}${x}_e^{(b)}$}  respectively. 
Normalized autocorrelation coefficients of respective signal are {\color{black}plotted in the bottom panel, and} to be discussed next.

\subsection{Subband Autocorrelation and Pitch Extraction}

The encoded subband signals are further processed via autocorrelation in frames of length $n_{corr} = \lceil 2\cdot f_s/F0_{min} \rceil$ and in step size of $n_{step} \in \mathbb{N} $. 
The range of sample delays is $d_{\tau} \in [d_{min}, d_{max}]$, where $d_{min} = \lfloor f_s/F0_{max} \rfloor $, $d_{max} = \lceil f_s/F0_{min} \rceil$. Here $\lfloor \cdot \rfloor$ denotes the largest integer less than or equal to a given number, while $\lceil \cdot \rceil$ denotes the smallest integer greater than or equal to a given number. 

Normalized autocorrelation coefficients (NAC) for encoded subband $b$ in the $j$th frame can be calculated using  
\begin{align} 
\label{eq:autocorr}
A^{(b)}(j,d_{\tau}) 
= \frac{ \sum_{k=(j-1)\cdot n_{step} + 1}^{(j-1)\cdot n_{step} + n_{corr} - d_{\tau}} \tilde{x}_{e}^{(b)}(k) \cdot \tilde{x}_e^{(b)}(k+d_{\tau}) }{
\sum_{k = (j-1)\cdot n_{step} + 1}^{(j-1)\cdot n_{step} + n_{corr} } [\tilde{x}_e^{(b)}(k)]^2 } ,
\end{align}
where 
\begin{equation} \label{eq:avgDCremove}
\tilde{x}_e^{(b)}(k) = {x}_e^{(b)}(k) - \frac{1}{n_{corr}} \cdot \sum_{k'=(j-1)\cdot n_{step} + 1}^{(j-1)\cdot n_{step} + n_{corr}} {x}_e^{(b)}(k')
.
\end{equation}
Compared with the cross-correlation function (see e.g. \cite{talkin1995robust}) for pitch estimation, this autocorrelation method results in a decreasing envelope as the time delay increases, due to the decreasing length of data in the numerator, which actually helps in suppressing the sub-harmonic errors. 
Similarly, the subband signal and the half-wave rectified subband signal can be used instead of the encoded subband signal in (\ref{eq:autocorr}) to calculate their corresponding NACs, and the results are given in the bottom panel of Fig.~\ref{fig:pitch_encoded}. 
We can see in this case that compared to the other two curves, the proposed subband encoding method produces a sharp peak corresponding to the expected period in the NAC, and there is no significant second peak in the expected range of sample delays $[d_{min}, d_{max}]$.

In each time frame, we use the average of the NAC over subbands:
\begin{equation}
\label{eq:sumAutoCorr}
A_{\sum}(j,d_{\tau}) = \frac{1}{N_b} \sum _{b=1}^{N_b} A^{(b)}(j,d_{\tau}) . 
\end{equation}

\begin{figure}[h]
\centering
\includegraphics[width=0.49\textwidth]{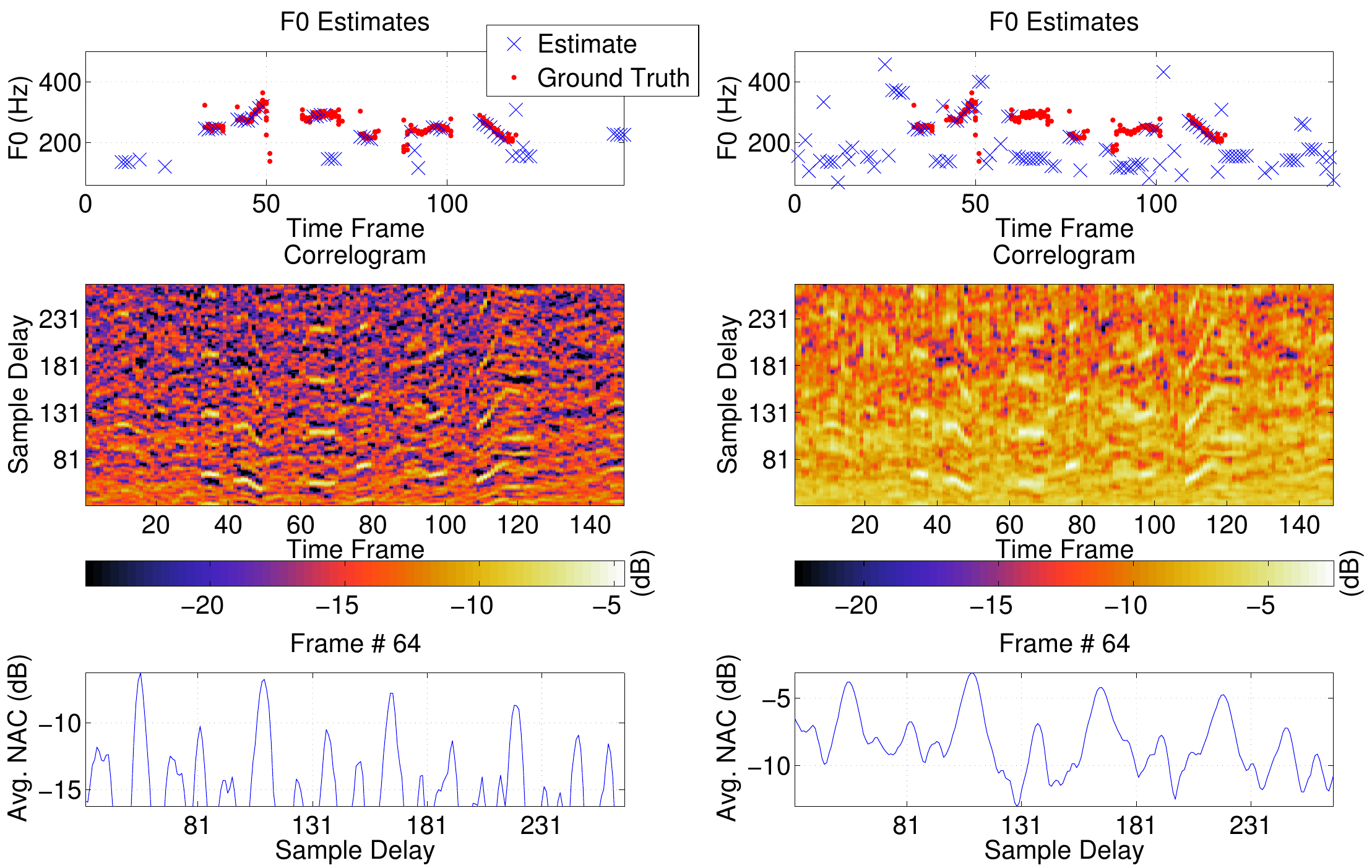}
\caption{Pitch estimation results (female speech with babble noise, SNR={\color{black}5}dB). Left column gives the pitch estimation results from proposed method. Right column shows the pitch estimation results using the autocorrelation of raw subband signals. 
}
\label{fig:pitch_autocorr}
\end{figure}

Then the pitch(es) in each frame can be estimated from the sample delays that correspond to the peaks in $A_{\sum}(j, \cdot)$. 
The strongest peak over the threshold $T_{A_{\sum}} = 0.125$ (i.e. $-9$dB) is directly used for the single pitch estimation {\color{black}(cf. the correlogram in Fig.~\ref{fig:pitch_autocorr} for the selection of this threshold, which is found consistent over a range of test cases)}. 
Due to the quasi-periodic nature of speech signals, for multi-pitch estimation, weaker peaks at sample delays that correspond to harmonics or sub-harmonics of the stronger peaks are removed. 
The pseudocode of the pitch extraction for frame $j$ is summarized in Algorithm~\ref{alg:pitchEstimator}. 
\begin{algorithm}
\caption{Pitch Extraction for the Pitch Estimator}
\label{alg:pitchEstimator}
\hspace*{\algorithmicindent} \textbf{Input:} normalized autocorrelation coefficients $A_{\sum}(j,\cdot)$; \\
\hspace*{\algorithmicindent} \textbf{Output:} pitch estimates $ \widehat{F_0} \{j\} $.
\begin{algorithmic}[1]
\Procedure{Pitch Extraction}{}
	\State \textbf{Find Peaks:}		
		\State Find and Sort all peaks over a threshold, i.e. $A_{\sum}(j,\hat{d}_{\tau_i}) \geq T_{A_{\sum}}, ~i=1,...,N_k$, and $A_{\sum}(j,\hat{d}_{\tau_1}) \geq A_{\sum}(j,\hat{d}_{\tau_2}) \geq \cdots $;
	\State \textbf{Pitch Extraction:}
		\If {$N_k == 0$}
			\State $\widehat{F_0} \{j\} = \emptyset$ (e.g. unvoiced sound segment or silence or miss-detection);
		\ElsIf {$N_k \geq 1$}
			\If {single pitch} 
			\begin{equation}
			\label{eq:pitchestset}
			\widehat{F_0} \{j\} =\{{f_s} / {\hat{d}_{\tau_1} }  \} .
			\end{equation}
			\ElsIf {multi-pitch}
			\For{$i=1:N_k$}
			\For{$\rho = i:N_k$}
				\If {$\hat{d}_{\tau_{\rho}}$ is a harmonic or sub-harmonic of $\hat{d}_{\tau_{i}}$, }
		Discard $\hat{d}_{\tau_{\rho}}$.
				\EndIf
			\EndFor
			\EndFor
			\State  $\widehat{F_0} \{j\} =\{\hat{f}_i ~|~ \hat{f}_i = {f_s} / {\hat{d}_{\tau_i} } ~\land~ \hat{f}_i \neq {f_s} / {\hat{d}_{\tau_{\rho}}} \} $.
		\EndIf
		\EndIf
\EndProcedure
\end{algorithmic}
\end{algorithm}

%

%
Fig.~\ref{fig:pitch_autocorr} provides a single pitch example comparing the proposed estimator (\ref{eq:sumAutoCorr}) using the encoded subband signals (\ref{eq:encodedSubband}) and using raw subband signals (\ref{eq:micSignal}). 
The top row provides the resulted pitch estimation results using (\ref{eq:pitchestset}) from the proposed method and the reference method. 
We can see that the proposed method produces more valid estimates, while the reference method produces considerably more errors. 
The middle row depicts the correlogram (\ref{eq:sumAutoCorr}) from the proposed method and the reference method. The proposed method produces more distinct pitch patterns. 
The bottom row shows the averaged autocorrelation results at frame 64, where the proposed method correctly produces the pitch estimate, while the reference method produces a sub-harmonic error. 
Therefore it is clear that the proposed method has distinct peaks by virtue of the proposed pitch encoding, while the peaks of the reference method are comparatively widespread. 
Moreover, using the raw subband signals produces more harmonics or sub-harmonics errors. 
For both cases, spurious estimates when there are no voiced sounds in the ground truth speech signal are from the babble noise.

As also can be seen from the top panel of Fig.~\ref{fig:pitch_autocorr}, the pitch estimates $ \widehat{F_0} \{ j \}$ from (\ref{eq:pitchestset}) contains in most cases the desired pitch estimates compared with the ground truth, but occasionally there may still be the sub-harmonics, harmonics or other spurious errors, which do not form continuous pitch contours with neighbouring estimates. It can also be an empty set, especially in the case of unvoiced sounds or silence.

\section{Pitch Tracker}
\label{sec:pitch-tracking}

In order to extract the desired pitch estimates of the speaker from $ \widehat{F_0} \{ j \}$, while suppressing the spurious errors (e.g. the pitch estimates that jump far away from pitch contours as shown in Fig.~\ref{fig:pitch_autocorr}), we exploit the temporal continuity constraint for pitch contour assuming continuous speech production by the human vocal system. 
Further assuming that the pitch of a speaker tend to return to its average level, we propose to model the pitch transition with the Ornstein Uhlenbeck process \cite{gardiner1985handbook}. 
For concurrent speakers, pitch tracking also aim at forming separate tracks of pitch estimates for respective speakers. However, more prior information (e.g. by training the algorithms) is usually required to separate pitches that overlap (i.e. when speakers have close levels of pitches). 
Nonetheless, we point out that in the case where pitches of concurrent speakers are at different levels, it is possible to track concurrent speaker pitches without the effort of training the algorithms. Thus we propose to treat the speaker as a target that has labeled (i.e. with identity) states (i.e. pitches) evolving over time, thereby tracking the pitch of speakers based on pitch observations (i.e. estimates $ \widehat{F_0} \{ j \}$ from Section \ref{sec:pitch-estimation}) and the GLMB\cite{vo2013labeled, vo2014labeled, reuter2014labeled} \footnote{
We briefly give necessary background on the GLMB in Subsections \ref{sec:GLMBrfsdef} to \ref{sec:GLMBprediction}. 
Readers are encouraged to refer to \cite{vo2013labeled, vo2014labeled, reuter2014labeled} and their references for detailed studies on GLMB, $\delta$-GLMB, LMB Bayes RFS tracking filters.} online tracking framework. 
We also implement a measurement driven birth model \cite{reuter2014labeled, lin2016measurement} for the practical adaptive pitch target births in the GLMB recursion, and present adaptations of the GLMB filter for the pitch tracking problem. 
%
In contrast to existing multi-pitch tracking methods \cite{wu2003multipitch, wohlmayr2011probabilistic}, the proposed method  does not require training, as the models used do not rely on the speech database.

\subsection{GLMB RFS Definitions}
\label{sec:GLMBrfsdef}

Denote the labeled state of pitch target as $\mathbf{x}_i \triangleq (\mathrm{x}_i, \ell_i) \in \mathbf{X} $, where $i$ is index, $\mathrm{x}_i$ denotes the pitch state, and $\ell_i$ its label (target identity).  
The GLMB RFS $\mathbf{X} \triangleq \{ (\mathrm{x}_i, \ell_i) ~|~ i \in \mathbb{N} \} $ is a labeled RFS \footnote{An RFS is a finite-set-valued random variable, whose number of points is random and the points are unordered and also random \cite{mahler2007statistical}.} with state space $\mathbb{X}, (\mathrm{x}_i \in \mathbb{X}) $ and label space $\mathbb{L}, (\ell_i \in \mathbb{L}) $, where the labels are unique, i.e. $\ell_{i} \neq \ell_{i'},~ \forall i \neq i'$. Its probability density is given as \cite{vo2013labeled} 
\begin{equation} \label{eq:GLMBRFS}
\mathbf{\pi }(\mathbf{X})=\Delta (\mathbf{X})\sum_{\xi \in \Xi }w^{(\xi )}(%
\mathcal{L(}\mathbf{X}))\left[ p^{(\xi )}\right] ^{\mathbf{X}} , 
\end{equation}
where the discrete index space $\Xi $ is the space of association map histories. 
Each $\xi \in \Xi$ represents a history of association map up to current time. 
Each $p^{(\xi )}(\cdot ,\ell )$ is the probability density of the states of target $\ell \in I=\mathcal{L(}\mathbf{X})$, and each $w^{(\xi )}(I)$ is non-negative with $%
\sum_{(I,\xi )\in \mathcal{F}\!(\mathbb{L})\!\times \!\Xi }w^{(\xi )}(I)=1$. 
Projection $\mathcal{L}:\mathbb{X}\mathcal{\times }\mathbb{L}\rightarrow \mathbb{L}$ is defined as $\mathcal{L}((\mathrm{x},\ell ))=\ell $, and $\mathcal{L}(\mathbf{X})=\{\mathcal{L}(\mathbf{x}) ~|~ \mathbf{x}\!\in \!\mathbf{X}\}$. 
$\mathbf{\mathcal{F}(}\mathbb{X)}$ denotes the class of finite subsets of a space $\mathbb{X}$.
The function $ \Delta (\mathbf{X}) \triangleq \delta _{|\mathbf{X}|} (|\mathcal{L}(\mathbf{X})|) $ is called the \emph{distinct label indicator}, where $|\cdot |$ denotes the cardinality of an RFS. 
The RFS exponential notation is defined as $h^{X} \triangleq \prod \nolimits _{\mathrm{x}\in X} h(\mathrm{x}) $, where $h$ is a real-valued function, with $h^{\emptyset }=1$ by convention.   
%

The $\delta$-GLMB form of (\ref{eq:GLMBRFS}) is completely characterized by the set of parameters $\{(\omega ^{(I,\xi )},p^{(\xi )}) ~|~ (I,\xi )\in \mathcal{F}\!(\mathbb{L})\!\times \!\Xi \}$, with the probability density given as \cite{vo2014labeled}
%
\begin{equation}
\mathbf{\pi }(\mathbf{X})=\Delta (\mathbf{X})
\!\!\!\! \sum_{(I,\xi )\in \mathcal{F}(%
\mathbb{L})\times \Xi }\omega ^{(I,\xi )}\delta _{I}(\mathcal{L(}\mathbf{X}))%
\left[ p^{(\xi )}\right] ^{\mathbf{X}} ,  \label{eq:generativeGLMB}
\end{equation}
where the pair $(I,\xi )\in \mathcal{F}(\mathbb{L})\times \Xi $ is called a
\emph{hypothesis}, and its associated weight $\omega ^{(I,\xi )}$ the probability of the hypothesis. 
\begin{equation*}
\delta _{Y}(X)\triangleq \left\{
\begin{array}{l}
1,\text{ if }X=Y \\
0,\text{ otherwise} . %
\end{array}%
\right. 
\end{equation*}%

The GLMB recursion consists of the ``update'' step based on Bayes inference and the Chapman-Kolmogorov \cite{gardiner1985handbook} ``prediction'' step based on the state transition model.

\subsection{GLMB Recursion: Update}
\label{sec:GLMBupdate}

If the current RFS prediction density is a $%
\delta $-GLMB of the form (\ref{eq:generativeGLMB}), using the current observation (pitch estimates) by denoting $\hat{F_0} \triangleq \widehat{F_0} \{ j \}$ as given in Algorithm \ref{alg:pitchEstimator}, the
posterior density is a $\delta $-GLMB \cite{vo2014labeled}, i.e.  
\allowdisplaybreaks
\begin{equation} 
\begin{aligned} 
 & \mathbf{\pi }\!(\mathbf{X}| \hat{F_0})= \\ &
\Delta \!(\mathbf{X})\!\!\!\!\!\!\!\!\sum_{(I,%
\xi )\in \mathcal{F}\!(\mathbb{L})\!\times \!\Xi }\;\sum\limits_{\theta \in \Theta \!(I)}\!\!\!\!\omega^{\!(I,\xi ,\theta \!)\!}(\hat{F_0}) \delta
_{\!I\!}(\mathcal{L\!(}\mathbf{X})\!)\!\!\left[ p^{\!(\xi ,\theta )\!}(\cdot
|\hat{F_0})\right] ^{\!\mathbf{X}} ,  \label{eq:PropBayes_strong0}
\end{aligned}%
\end{equation}
where $\Theta (I)$ denotes the subset of current association maps with
domain $I$,\allowdisplaybreaks%
\begin{eqnarray}
\omega ^{(I,\xi ,\theta )\!}(\hat{F_0})\!\!\! &\propto &\!\!\!\omega ^{(I,\xi
)}[\eta _{F_0}^{(\xi ,\theta )}]^{I}  \label{eq:PropBayes_strong1} \\
p^{\!(\xi ,\theta )\!}(\mathrm{x},\ell |\hat{F_0})\!\!\! &=&\!\!\!\frac{p^{(\xi )}(\mathrm{x},\ell
)\psi _{F_0}(\mathrm{x},\ell ;\theta )}{\eta _{F_0}^{(\xi ,\theta )}(\ell )}
\label{eq:PropBayes_strong3} \\
\psi _{F_0}(\mathrm{x},\ell ;\theta )&=&\!\!\!\!\! \left\{
\begin{array}{ll}
\!\!\!\! \frac{p_{D}(\mathrm{x},\ell ) \mathrm{g} (\hat{f}_{\theta (\ell )}|\mathrm{x},\ell )}{\kappa (\hat{f}_{\theta (\ell
)})}, \text{if }\theta (\ell )>0 \\
1-p_{D}(\mathrm{x},\ell ), \text{if }\theta (\ell )=0%
\end{array}%
\right.  \label{eq:PropConj5}  \\
\eta _{F_0}^{(\xi ,\theta )}(\ell )\!\!\! &=&\!\!\!\left\langle p^{(\xi
)}(\cdot ,\ell ),\psi _{F_0}(\cdot ,\ell ;\theta )\right\rangle 
\label{eq:PropBayes_strong2}
\end{eqnarray}
$\mathrm{g}(\hat{f}_{\theta (\ell )}|\mathrm{x},\ell )$ is the likelihood for the measurement $\hat{f}_{\theta (\ell )} \in \hat{F_0} $ being generated by $(\mathrm{x},\ell)$, and $\kappa(\cdot)$ is the intensity function of Poisson RFS which we use to describe the clutter. $p_D$ is the probability of a target state being detected. 
The standard inner product notation is defined as $\left\langle f,g\right\rangle \triangleq \int f(\mathrm{x})g(\mathrm{x})d\mathrm{x} $.

\subsection{GLMB Recursion: Prediction}
\label{sec:GLMBprediction}

If the current RFS filtering density from its previous update step is a $%
\delta $-GLMB of the form (\ref{eq:generativeGLMB}), the prediction density to the next time is a $\delta $-GLMB given as \cite{vo2014labeled}
\begin{equation}
\begin{aligned}
\mathbf{\pi }_{\! +} & (\mathbf{X}_{\!+\!})
=  \\ & \Delta(\mathbf{X}%
_{\!+})\!\!\!\!\!\!\!\sum_{(I_{+},\xi )\in \mathcal{F}(\mathbb{L}_{+})\times
\Xi }\!\!\!\!\omega _{+}^{(I_{+},\xi )}\delta _{I_{+\!}}(\mathcal{L(}\mathbf{X}%
_{\!+}))\!\left[ p_{+}^{(\xi )\!}\right] ^{\!\mathbf{X}_{+}} , 
\label{eq:PropCKstrong1}
\end{aligned}%
\end{equation}
%
%
%
where\allowdisplaybreaks%
\begin{eqnarray}
\!\!\!\omega_+ ^{(I_+,\xi )}\!\! &=&\!\!\omega _{S}^{(\xi )}(I_{+}\cap
\mathbb{L}) w_{B}(I_{+}\cap \mathbb{B})  \label{eq:PropCKstrong2} \\
\!\!\!p_{+}^{(\xi )}(\mathrm{x},\ell )\!\! &=&\!\!1_{\mathbb{L}}(\ell )p_{S}^{(\xi
)\!}(\mathrm{x},\ell )+1_{\mathbb{B}\!}(\ell )p_{B}(\mathrm{x},\ell )  \label{eq:PropCKstrong3}
\\
\!\!\!\omega _{S}^{(\xi )}(L)\!\! &=&\!\![\eta _{S}^{(\xi
)}]^{L}\sum_{I\supseteq L}[1-\eta _{S}^{(\xi )}]^{I-L}\omega ^{(I,\xi )}
\label{eq:PropCKstrongws} \\
\!\!\!p_{S}^{(\xi )}(\mathrm{x},\ell )\!\! &=&\!\!\frac{\left\langle p_{S}(\cdot
,\ell ) \mathrm{f}(\mathrm{x}|\cdot ,\ell ),p^{(\xi )}(\cdot ,\ell )\right\rangle }{\eta
_{S}^{(\xi )}(\ell )}  \label{eq:PropCKstrong4} \\
\!\!\!\eta _{S}^{(\xi )}(\ell )\!\! &=&\!\!\left\langle p_{S}(\cdot ,\ell
),p^{(\xi )}(\cdot ,\ell )\right\rangle   \label{eq:PropCKstrong_eta} 
\end{eqnarray}
$[\cdot]_+$ stands for prediction, $ \mathrm{f}(\mathrm{x}|\cdot ,\ell )$ is the state transition function that we will propose in Section \ref{sec:pitchtransition}. $\mathbb{B}$ is the space of new-born target labels. The set of new-born targets can be represented by an LMB RFS, where $w_{B}$ is the probability of a birth hypothesis of new-born targets 
and $p_B$ is the probability distribution of pitch states that belong to the birth targets as will be detailed in Section \ref{sec:MDB}. $p_S(\cdot, \ell)$ is the survival probability.
The inclusion function, a generalization of the indicator function is defined as %
\begin{equation}
1_{Y}(X)\triangleq \left\{
\begin{array}{l}
1,\text{ if }X\subseteq Y \\
0,\text{ otherwise} . %
\end{array}%
\right. 
\end{equation}%


\subsection{The Pitch Transition Model}
\label{sec:pitchtransition}

A possible way of exploiting the temporal continuity for pitch tracking is using the hidden Markov model (HMM) \cite{wu2003multipitch, lee2012noise, wang2017robust}. While it is reasonable and useful to model the pitch transition as a Markov process, the HMM however, usually requires training the algorithms to obtain \textit{a priori} knowledge of the state transition probabilities, which can be inconvenient and restricting in practice. 

It is well-known that the pitch of a human speaker depends on the vocal tract, sub-glottal resonance and speech content. Because of the pronunciation, intonation and emotion, the fundamental frequency of a human speaker can vary over a continuous range \cite{deller1993discrete}.
This range is usually a limited subset of $[F0_{min}, F0_{max}]$, and naturally, the pitch of a human speaker tend to move toward its average level over time. 

Therefore, assuming a Gaussian distribution of pitch states centered at the speaker's mean pitch value over time, and using the temporal continuity constraint, we propose to model the speaker pitch transition function $ \mathrm{f}(\mathrm{x}_{\hat{f}} | \hat{f} ,\ell ) $ in (\ref{eq:PropCKstrong4}) as an Ornstein-Uhlenbeck process \cite{gardiner1985handbook}, i.e. 
\begin{equation}
\label{eq:mean-reversion}
\mathrm{x}_{\hat{f}} = \hat{f} + \alpha \cdot (\mu_{\hat{q}} - \hat{f} ) \cdot t_{step} + \nu_{\sigma_{\hat{q}}} , 
\end{equation}
%
%
where $\hat{f} \in \hat{F_0} $ denotes a measurement of pitch (from the pitch estimator) at current time, and $\mathrm{x}_{\hat{f}}$ denotes the ``predicted'' pitch state at next time frame. Parameters $\mu_{\hat{q}} > 0$ and $\sigma_{\hat{q}} >0 $ are respectively the mean value and standard deviation of the pitch with index ${\hat{q}}$. The reversion rate $\alpha > 0$ specifies how fast the pitch return to its mean, and $t_{step}$ ($t_{step} = n_{step} / f_s $) is the time step. We choose $\alpha=0.1$ in this paper. $\nu_{\sigma_{\hat{q}}} \sim \mathcal{N}(0,\sigma_{\hat{q}})$ is the Gaussian distribution with mean value of $0$ and standard deviation of $\sigma_{\hat{q}}$. 
Apparently (\ref{eq:mean-reversion}) describes a Gaussian and Markov process \cite{gardiner1985handbook} with long term mean of $\mu_{\hat{q}}$, hence is also called a mean-reverting process.

In (\ref{eq:mean-reversion}), an arbitrary pitch measurement $\hat{f} \in [F0_{min}, F0_{max}] $ is mapped to an index ${\hat{q}}$ via: 
\begin{equation} \label{eq:q}
{\hat{q}} = \argmin_q | \hat{f} - \mu_q | ,~ \mu_q \in \vec{\mu} , 
\end{equation} 
where $\mu_q$ span the pitch range $[F0_{min}, F0_{max}]$ evenly with steps of $\mu_S > 0$, i.e. 
\begin{equation} \label{eq:vecq}
\vec{\mu} = \{ \mu_q ~|~ \mu_q = F0_{min}+ \mu_S \cdot (q - 1/2),~ q = 1,..., q_M  \} , 
\end{equation}
where $q_M = \lfloor (F0_{max}-F0_{min})/\mu_S \rfloor $. This is reasonable since we have no \textit{a priori} knowledge of the pitch level,  sampling the pitch range with $\mu_q$ initializes the mean-reverting process. 
%

Typically a greater $\sigma_q$ corresponds to a greater $\mu_q$. Thus with a first-order approximation of coefficient $\kappa_{\mu} \in (0,1) $, we have
\begin{equation} \label{eq:sigmaq}
\sigma_q = \kappa_{\mu} \cdot \mu_q .
\end{equation}

With a step size $\mu_S$ not too large, and the coefficient $\kappa_{\mu}$ not too small, reasonable sampling of pitch range can be obtained. 
We choose $\mu_S=40$Hz and $\kappa_{\mu}=0.1$ in this paper.

\subsection{Measurement Driven Birth} \label{sec:MDB}

The standard implementation of GLMB filter in (\ref{eq:PropCKstrong2}) and (\ref{eq:PropCKstrong3}) in Section \ref{sec:GLMBprediction} relies on \textit{a priori} knowledge of target birth distributions, which restricts its applications in practice. Here we adapt the measurement-driven birth model that we presented in \cite{lin2016measurement} for pitch tracking. It initiates the pitch states and existence probabilities of birth targets based on measurement data (pitch estimates) from previous time, hence adaptively tracks speaker pitches online. More details of measurement driven birth model for LMB and GLMB can be found in \cite{reuter2014labeled, lin2016measurement} respectively.

In the GLMB update step, measurements $\hat{f} \in \hat{F_0}$ are associated with persistent tracks and the corresponding hypothesis probability $\omega ^{(I,\xi ,\theta )\!}(\hat{F_0})$ in (\ref{eq:PropBayes_strong1}) as well as the probability density $p^{\!(\xi ,\theta )\!}(\mathrm{x},\ell |\hat{F_0}) $ in (\ref{eq:PropBayes_strong3}) are calculated. 
According to the corresponding hypothesis probability, each pitch measurement $\hat{f}$ initiates new-born targets at the next time step, with the new-born likelihood for each measurement $\hat{f} \in \hat{F}_0$ found by
\begin{equation} \label{eq:rU}
r_{N} (\hat{f}) = 1 - \sum_{(I,\xi )\in \mathcal{F}\!(\mathbb{L})\!\times \!\Xi } ~ \sum\limits_{\theta \in \Theta \!(I)} 1_{\hat{f}_{\theta}}(\hat{f}) \omega^{(I,\xi,\theta)} ,
\end{equation}
where the inclusion function indicates if the measurement $\hat{f}$ has been assigned to a target by any of the updated hypotheses. It can be seen from (\ref{eq:rU}) that, a measurement that has been used in all hypotheses cannot initiate a new-born target ($r_{N} (\hat{f}) =0$), while for measurements that have not been assigned to any of the targets, the new-born likelihood is 1.

For each measurement $\hat{f}$ that has non-zero new-born likelihood, a new birth of Bernoulli RFS is generated around the measurement, assuming a Gaussian distribution. Thus the probability distribution of the states $p_{B}(\mathrm{x},\ell )$ in (\ref{eq:PropCKstrong3}) for the measurement-driven birth model is given as,
\begin{equation}
\label{eq:MDBdensity}
p_{B}(\mathrm{x},\ell;\hat{f}) = \sum_{i=1}^{M_b} \frac{1}{M_b} \delta_{\mathrm{x}_{\hat{f}}^{(i)}}(\mathrm{x}), \; \hat{f} \in \hat{F}_0
\end{equation} 
\begin{equation} \label{eq:gaussBirth}
\mathrm{x}_{\hat{f}}^{(i)} \sim \mathcal{N}%
\big( m_B(\hat{f}),P_B(\hat{f}) \big),\; i=1,...,M_b
\end{equation}
%
%
where $M_b$ denotes the number of generated states for the birth target. $m_B(\hat{f}) = \mu_{q_B}$ where $q_B$ is found from (\ref{eq:q}). $P_B(\hat{f}) = \sigma_{q_B}$ is a variance that specifies the distribution of states of the new-born target, and can be found from (\ref{eq:sigmaq}). Larger values of $P_B(\hat{f})$ result in higher error tolerance, while smaller values give better accuracy in general.

Thus the set of new-born targets is a labeled multi-Bernoulli RFS with the probability density given as \cite{reuter2014labeled} 
\begin{equation}
\mathbf{\pi } _{B} (\mathbf{X}_{+})
=   \Delta(\mathbf{X}%
_{+\!}) w_{B}(\mathcal{L}(\mathbf{X}_{+})) \left[ p_{B}\!\right] ^{\!\mathbf{X}_{+\!}} , 
\label{eq:lmbBirth}
\end{equation}
where the birth probability also required in (\ref{eq:PropCKstrong2}) is 
\begin{equation}
\label{eq:omegaB}
w _{B}(I) = \prod\limits_{i\in \mathbb{B}}\left( 1-r_{B}^{(i)}\right)
\prod\limits_{\ell \in I}\frac{1_{\mathbb{B}}(\ell )r_{B}^{(\ell )}}{1-r_{B}^{(\ell
)}} , 
\end{equation}
and the existence probability of the Bernoulli MDB at the next time that is initiated by a measurement $\hat{f} \in \hat{F}_0 $ depends on its new-born likelihood obtained from current time:
\begin{equation}
r_{B}(\hat{f}) = \min \Big( r_{B_{\max}},\;   \lambda_{B} \cdot  \frac{r_{N}(\hat{f})}{\sum_{\zeta \in \hat{F}_0 } r_{N}(\zeta) }  \Big) , 
\end{equation}
where $\lambda_{B}$ is the expected number of target birth at the next time, and $r_{B_{\max}} \in (0,1]$ is the maximum existence probability of a new-born target to ensure that the resulting $r_{B}(\hat{f})$ does not exceed 1 when $\lambda_{B}$ is too large. 
%
We choose $M_b = 1000$, $\lambda_{B}=0.3$ and $r_{B_{\max}}=0.15$ in this paper as in \cite{lin2016measurement}.

\subsection{Adaptations for Pitch Tracks}
\label{sec:adapt}

The above implementation of the GLMB filter can produce the number of pitch targets and the estimates of target pitch with respective labels (identities) over time. However, when pitch tracks are well apart in time, they tend to be assigned with different target identities, even if they are from the same speaker. Moreover, for the single-pitch tracking, it may still produce two or more targets due to spurious errors. Thus we propose further adaptations here for these two cases.

\subsubsection{Labeling Adaptation}
In practice, it is common that a same speaker can have pauses during speech thus should be assigned with a same label. 

Assuming that pitches of a close level belong to one target, we provide an adaptation for target labeling here. 
Once a new pitch target is confirmed, we compare its pitch estimate with all previously confirmed tracks. If the smallest difference is less than $20\%$ of the mean value of a particular pitch track, we assign the label of that track to the new target and update the association map by marking the pause or unvoiced periods as miss detections. 

%
%

\subsubsection{Single Pitch Extraction}

The single pitch extraction adaptation is {\color{black}proposed} by selecting the pitch target with the highest accumulated probability from all hypotheses. \myedit{ It is used when there is only one speaker unless otherwise noted.}

The pitch estimate is found with the label  
\begin{equation}
\hat{\ell} = \argmax_{\ell} w(\ell) , 
\end{equation}
where {\color{black}directly from the definition in (\ref{eq:PropBayes_strong0}), the accumulated probability of all hypotheses containing label $\ell$ is found, i.e.}
\begin{equation}
w(\ell) = \sum_{I \ni \ell} \omega ^{(I,\xi ,\theta )\!}(\hat{F_0}) . 
\end{equation}
%


\section{Numerical Studies}
\label{sec:performance}
This section demonstrates the performance of the proposed pitch estimator and tracker under various conditions. 
We first provide the pitch estimation and tracking results for the case of a single speaker, in presence of additive noises at various levels of signal to noise ratios (SNR). 
Then we also provide multi-pitch estimation and tracking results for concurrent speakers. 
The sound corpora used are from the CSTR database \cite{bagshaw1993enhanced,fdaSpeech}, which include 50 English {\color{black}utterances} from a male and a female speaker respectively and their {\color{black}laryngograph signals}. 
{\color{black}The Keele database is also used for verification \cite{plante1995pitch}.
}
The noise signals used {\color{black}include the white Gaussian noise as well as those} from the AURORA database \cite{varga1993assessment,auroraNoise}, which are composed of 8 types of noises from different environments.

\subsection{Experimental Setup}

All sound signals are resampled first at $f_s = 16000$Hz. 
For the pitch tracker, the detection probability for the pitch estimator is modeled as 
$p_D(\mathrm{x}, \cdot) \sim  p_{D_{max}} \cdot \mathcal{N}(\mathrm{x};f_{mid},R^2) $, where $p_{D_{max}} = 0.98$, $f_{mid} = \frac{1}{2} \cdot (F0_{min} + F0_{max})$ and $R = 100$Hz. 
The measurement likelihood for the GLMB update is $g(\hat{f}|x) \sim  \mathcal{N}(\hat{f};x,D)$, where ${D}= 5$Hz. 
The survival probability is $p_S = 0.8$, and the clutter rate is $\kappa = 0.0001$. 
%
%
%
The single speaker adaptation is applied for the case of single speaker, while the labeling adaptation is used for the case of multi-pitch tracker for concurrent speakers.
%

\subsection{Performance Metric{\color{black}s}}

We use the standard gross pitch error (GPE) for evaluating the {\color{black}accuracy of pitch estimates in voiced regions} \cite{chu2012safe, wang2017robust}.
\begin{equation}
\mathrm{GPE} {\color{black}\triangleq} \frac{N_{err}}{N_v} , 
\end{equation}
where $N_{err}$ is the number of frames with pitch estimates that deviate from ground truth by more than $5\%$, and $N_v$ denotes the total number of voiced frames as reported by both the ground truth and the estimation method. 

{\color{black}The voicing decision error (VDE) metric is also used to  evaluate the accuracy in deciding voiced/unvoiced frames, i.e.
\begin{equation}
\mathrm{VDE} \triangleq \frac{N_{ue}+ N_{ve}}{N_f} , 
\end{equation}
where $N_{ue}$ is the number of frames that have pitch estimates but are unvoiced from ground truth, $N_{ve}$ is the number of frames that have no pitch estimates but are actually voiced, and $N_f$ is the total number of frames.\footnote{\color{black}Note that for estimators that produce a pitch estimate in every frame (e.g. PEFAC or YIN), $N_{ve} \equiv 0$, and the resulting VDE may be biased toward the ratio of the total number of unvoiced frames to ${N_f}$, if the respective voicing decision measure (e.g. voice probability or  aperiodicity) is not used. 
}
} 

\myedit{
For multi-pitch tracking, the GPE or VDE measures may not suffice. Thus we evaluate the performance also with the speaker identity error (SIE), i.e.
\begin{equation} \label{eq:SIE}
\mathrm{SIE}_i \triangleq \frac{E_{ij}}{N_{v_i}} ,~ i,j~\in~ \{1,2,\cdots\},  
\end{equation}
where $N_{v_i}$ denotes the number of voiced estimates for speaker $i$ reported by both the ground truth and the pitch tracker, and $E_{ij}$ denotes the number of pitch estimates that are assigned to speaker $i$, but actually belong to speaker $j$. 
}

\subsection{Single Speaker}

\begin{figure}[!ht]
\centering
\includegraphics[width=0.49\textwidth]{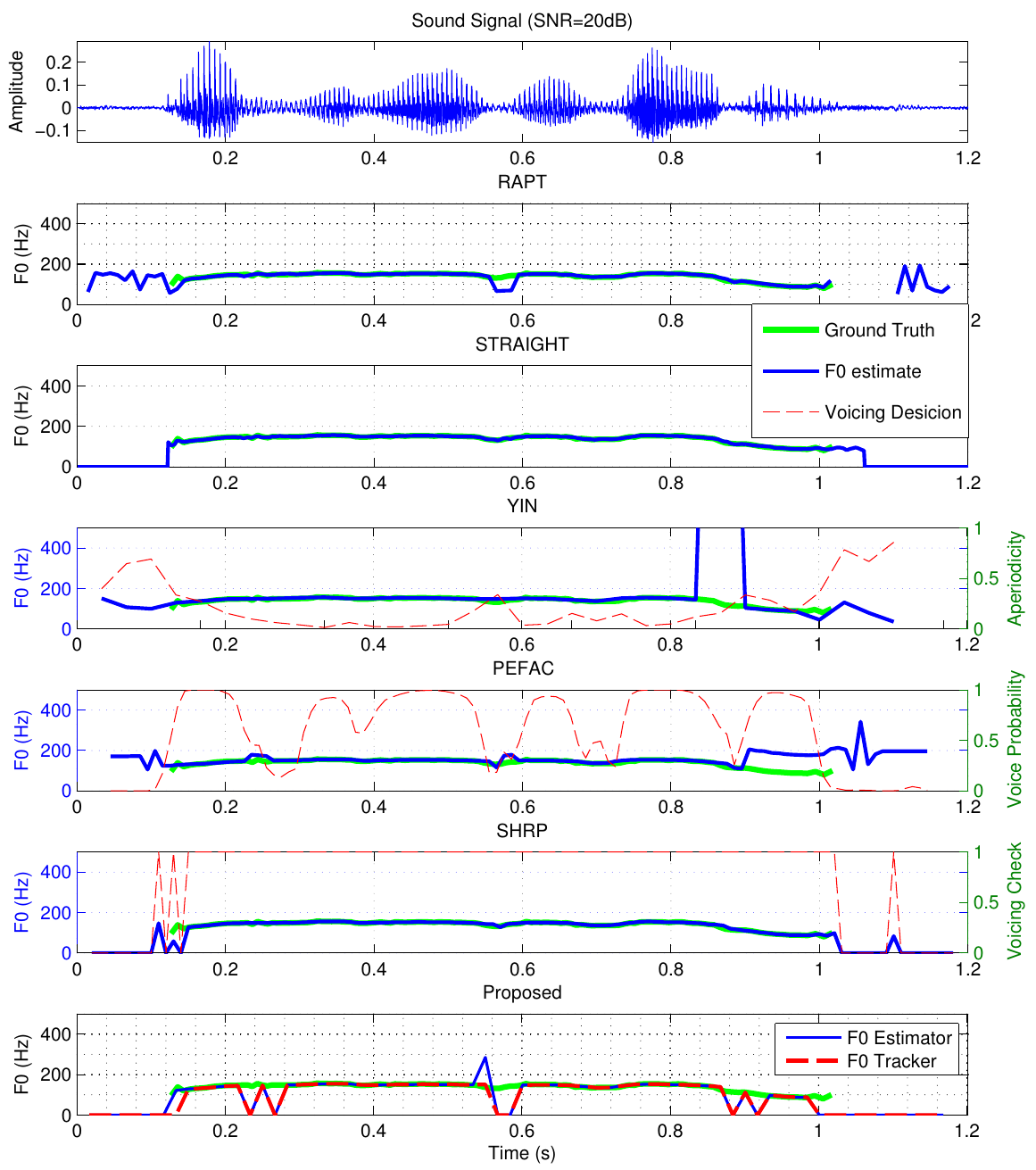}
\caption{From top to bottom: waveform of the speech signal with babble noise at SNR of 20dB (top panel), pitch ground truth of clean speech signal and pitch estimation results from the RAPT, \myedit{STRAIGHT,} YIN, PEFAC, SHRP and the proposed methods, respectively. }
\label{fig:singlespeakerPitch1}
\end{figure}
\begin{figure}[!ht]
\centering
\includegraphics[width=0.49\textwidth]{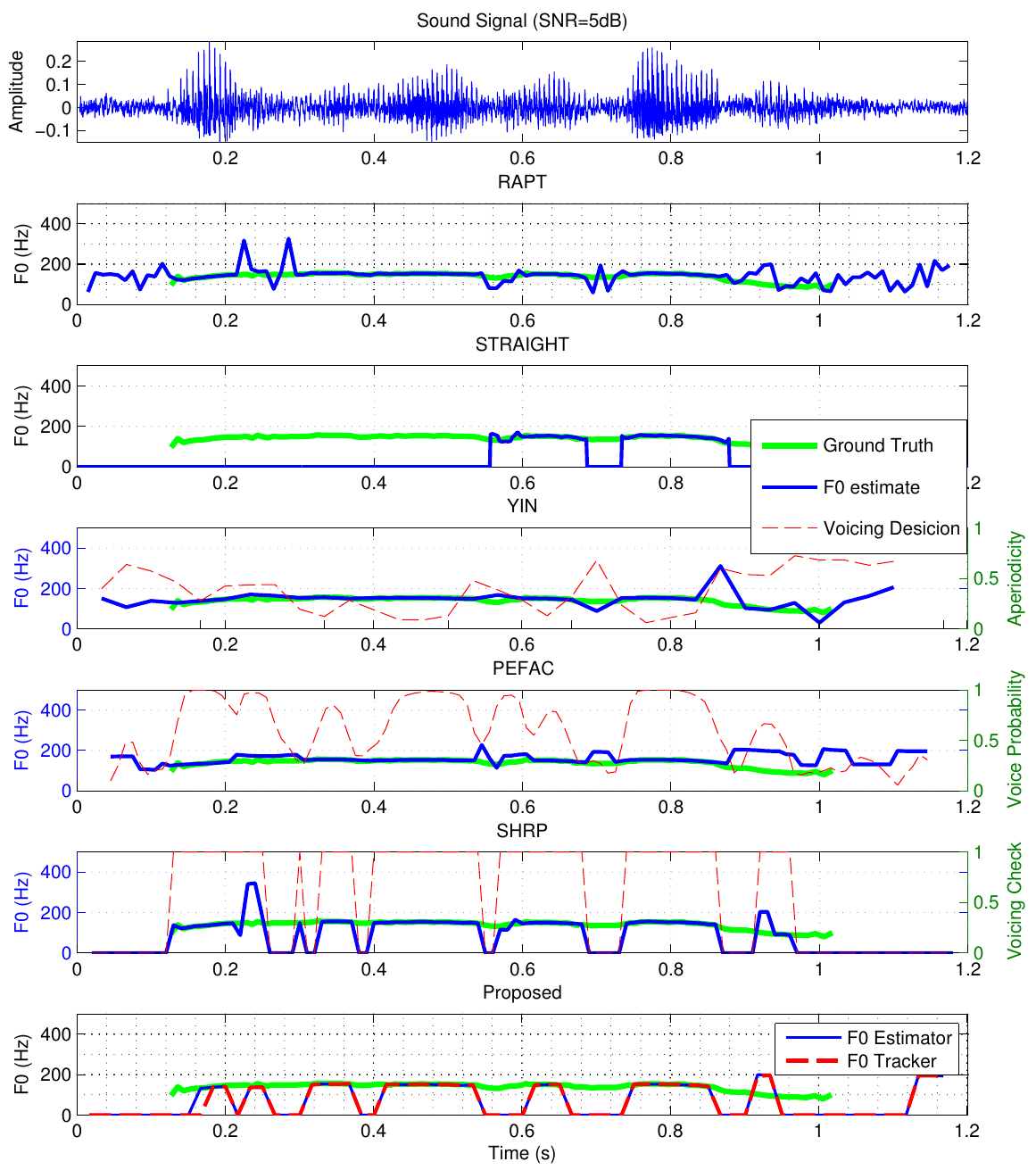}
\caption{From top to bottom: waveform of the speech signal with babble noise at SNR of {\color{black}5}dB (top panel), pitch ground truth of clean speech signal and pitch estimation results from the RAPT, \myedit{STRAIGHT,} YIN, PEFAC, SHRP and the proposed methods, respectively. }
\label{fig:singlespeakerPitch2}
\end{figure}

Fig.~\ref{fig:singlespeakerPitch1} shows the speech signal and pitch estimation results for a male speaker with additive babble noise, using the RAPT {\color{black}(Robust Algorithm for Pitch Tracking)} \cite{talkin1995robust}, YIN \cite{de2002yin}, PEFAC {\color{black}(Pitch Estimation Filter with Amplitude Compression)} \cite{gonzalez2014pefac}, SHRP {\color{black}(Subharmonic-to-Harmonic Ratio based Pitch determination algorithm)} \cite{sun2002pitch}, {\color{black}STRAIGHT (Speech Transformation and Representation using Adaptive Interpolation of weighted spectrum)} and our proposed methods, at SNR of 2{\color{black}0}dB. 
Pitch ground truth is plotted {\color{black}in green} as reference for each method. 
This figure represents the cases when the noise is weak. All the methods can produce accurate pitch estimates during the voiced period for most of the time, compared to the ground truth. 
The RAPT method produces spurious estimates at about $0.56$s, and also for the babble noise before about $0.1$s and after about $1.1$s. 
{\color{black}The STRAIGHT produces perfect estimates in this case, except the tail after about 1s. }
The YIN method uses the ``Aperiodicity'' measure as a voiced/unvoiced sound detector, however, a low aperiodicity can also correspond to erroneous estimates at around $0.85$s. 
The PEFAC method provides ``Voice Probability'' for detection of voiced/unvoiced sounds. The SHRP method provides a binary ``Voicing Check'' measure. The ``Aperiodicity'', ``Voice Probability'' and ``Voicing Check'' can be used as the voiced/unvoiced activity detector (VAD) for respective methods{\color{black}, and are plotted in red}.
Both the PEFAC and SHRP methods are frequency domain methods, and their estimates are regarded reliable when the corresponding {\color{black}voicing decision} measures are high (i.e. ``Voice Probability'' is close to 1 or the ``Voicing Check'' equals 1). 
However, {\color{black}we can also see that these voicing decision measures may also have outliers.
For example, the PEFAC produces correct estimates at time of about $0.27$s and $0.85$s while its voicing decision measure is close to 0. The SHRP produces inaccurate estimates at about $0.15$s and after $1$s where its voicing decision measure is 1. } 
The proposed pitch estimator and {\color{black}single} pitch tracker produces comparatively reliable results. The pitch estimates from the proposed pitch estimator and pitch tracker are all close to ground truth. There are miss-detections (i.e. empty set of estimates, plotted as zeros for clarity) at about $0.25$s, $0.55$s and $0.9$s, which correspond to the time segments when the voiced speech signal is weak. In this case, our proposed method produces no estimate for the time period dominated by weak (SNR=20dB) babble noise (i.e. before about $0.1$s and after about $1.1$s). 
Note that there is a spurious estimate at about 0.55s, and the pitch tracker successfully filters it.\footnote{\myedit{Note also that although the pitch tracker is useful at higher SNRs (e.g. SNR$\geq0$dB), it may not be able to improve the pitch estimation performance at very low SNRs (e.g. SNR$\leq-5$dB) due to excessive spurious estimates and miss-detections.}}
When there is no spurious estimate from the pitch estimator, the pitch estimates from the proposed pitch tracker almost overlap with those from the pitch estimator. The adaptive measurement driven birth model of the pitch tracker requires the initial measurements before confirming a new track, which can be seen at time of about $0.15$s.

\begin{figure*}[t]
\centering
\includegraphics[width=0.98\textwidth]{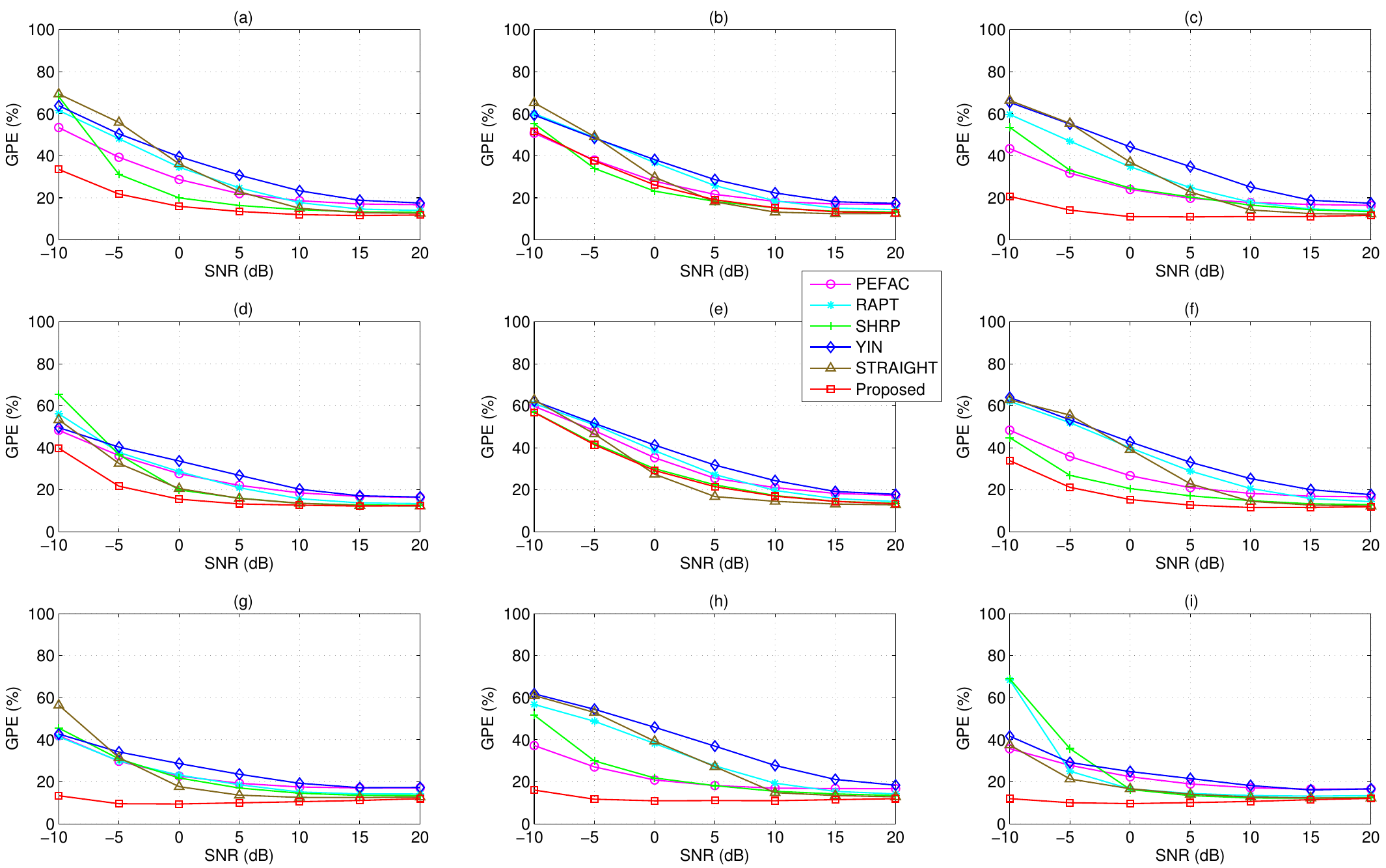}
\caption{\myedit{Averaged} GPE results for pitch estimation. Speech signals are from the CSTR and \myedit{Keele corpora}, while the additive noises are from the Aurora database and AWGN. Noise types are respectively (a) airport, (b) babble, (c) car, (d) exhibition, (e) restaurant, (f) street, (g) subway, (h) train, (i) AWGN.}
\label{fig:pitch_GPE}
\end{figure*}
Fig.~\ref{fig:singlespeakerPitch2} shows the case when the additive noise is strong (SNR= $5$dB). 
All existing methods produce spurious pitch estimates during the voiced period, compared with the ground truth. 
The RAPT produces considerable errors over time, which can be hard to suppress. 
{\color{black}The STRAIGHT produces correct pitch estimates between about 0.55s and 0.9s.}
The other methods can take advantage of their corresponding voice {\color{black}decision} measures. 
The YIN uses the aperiodicity as {\color{black}its voicing decision} measure, i.e. the estimates can be removed when the aperiodicity is high. 
Comparing Fig.~\ref{fig:singlespeakerPitch1} and Fig.~\ref{fig:singlespeakerPitch2}, in this paper, any estimate corresponding to an aperiodicity of greater than $0.5$ is discarded. 
For PEFAC, we choose the estimates with a voice probability of no less than $0.5$.
For SHRP, all estimates corresponding to voice check of $1$ are used. 
Our proposed pitch estimator and tracker can produce accurate pitch estimates during the voice period, although having more miss-detections compared with that of Fig.~\ref{fig:singlespeakerPitch1} due to the stronger babble noise. All estimates from our proposed methods are used for further quantitative comparison using the GPE metric. 
Note that by applying the VADs of respective reference methods, the GPE evaluates the accuracy of all selected pitch estimates, without taking into account those discarded ones. 
In general, all the {\color{black}five} state-of-the-art methods (RAPT, YIN, PEFAC{\color{black}, STRAIGHT} and SHRP) can provide reasonably reliable pitch estimates for high SNR sounds, but show different levels of performance degradation as the SNR drops. 
The spurious estimates can be suppressed to different extents, using their corresponding VADs. 
Overall, our proposed pitch estimator has produced reliable pitch estimates for voiced speech. For the sake of clarity in Fig.~\ref{fig:singlespeakerPitch1} and Fig.~\ref{fig:singlespeakerPitch2}, when the estimate is an empty set, we plot the pitch value as a zero. Similar to the reference methods, discarded pitch estimates (empty sets of pitch estimates in our proposed methods) are not counted in the GPE measure. 

%
%
In Fig.~\ref{fig:pitch_GPE}, we show the GPE results for all the \myedit{pitch estimators} using the CSTR \myedit{and the Keele corpora} with various types of noise and SNR levels. \myedit{The GPEs are averaged over all sound files.}
It may be arguable as to the fair selection of parameter values for best performance of respective methods. Here the parameters for RAPT, YIN, STRAIGHT, PEFAC and SHRP all use the default values as provided in respective programs, and in particular, frame lengths are $10$ms, $33.3$ms, $80$ms, $10$ms and $10$ms respectively. 
We choose a frame length of $33.3$ms for our proposed method, which is two periods of the minimum F0 frequency ($F0_{min} = 60$Hz). 
%
We can see from Fig.~\ref{fig:pitch_GPE} that all methods degrade as the noise get stronger. However, the proposed method outperforms the other state-of-the-art methods in most cases. The performance of the proposed method is worst at the babble noise or the restaurant noise, both of which are basically random mixtures of human speech signals. 
It is also interesting to notice that the STRAIGHT method performs better than most other methods at high SNRs.
Moreover, compared with other noise types, the additive white Gaussian noise (AWGN) seems to cause least degradation to all these pitch estimators, except the outliers from the RAPT and SHRP at SNR$\leq\!\!-5$dB.

\begin{figure}[!ht]
\centering
\includegraphics[width=0.49\textwidth]{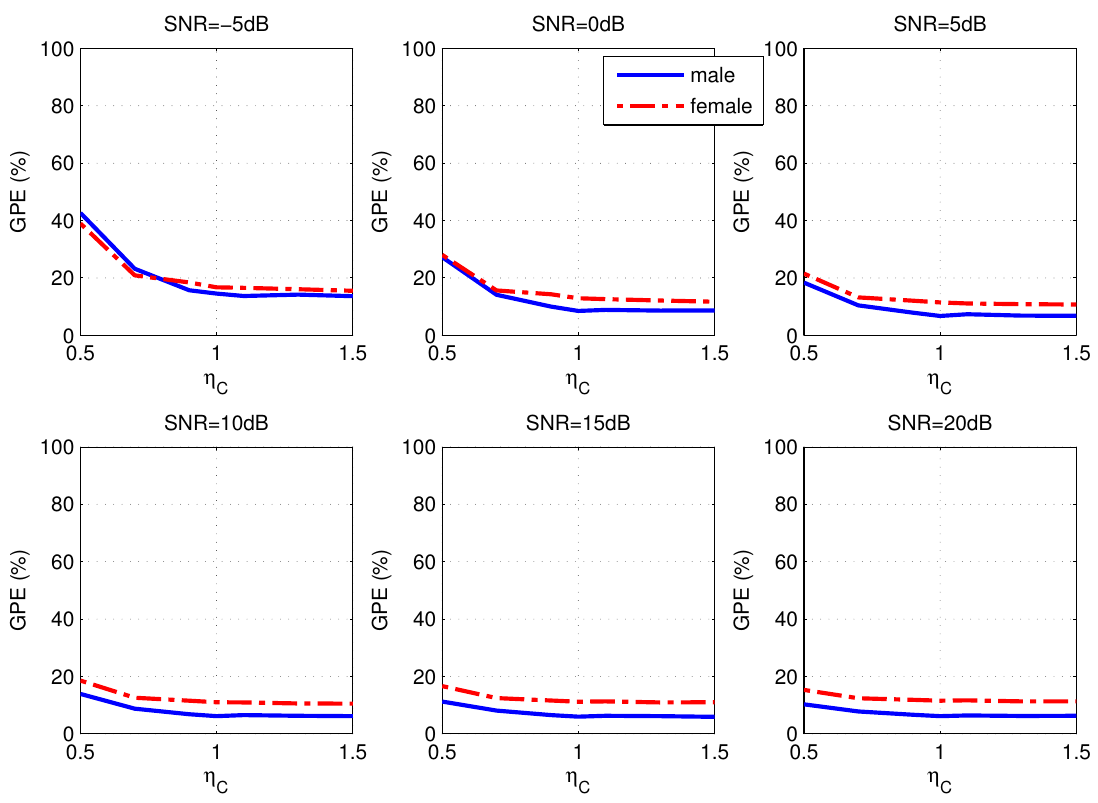}
\caption{GPE versus frequency coverage $\eta_C$ using pitch estimation results (averaged over all noise types and sound files). }
\label{fig:gpe_etaC}
\end{figure}

The proposed frequency coverage is verified in Fig.~\ref{fig:gpe_etaC} where we test and check the GPE results for the male and female speakers of the CSTR corpus at various frequency coverage. Here the GPE is an averaged result \myedit{from the pitch estimator} over all noise types. We can clearly see that despite the changes of SNR, the accuracy improves (the gross pitch error decreases) as $\eta_C$ increases until $\eta_C = 1$, and the GPE is comparatively stable for $\eta_C \in [1, 1.5]$. We know that as $\eta_C$ increases, the number of subbands also increases, thus requiring more computations. Hence as we have expected in Section \ref{sec: freqCov}, $\eta_C = 1$ is chosen for good estimation accuracy and low computational load for our proposed methods. 
%

%

\begin{figure}[!h]
\centering
\includegraphics[width=0.49\textwidth]{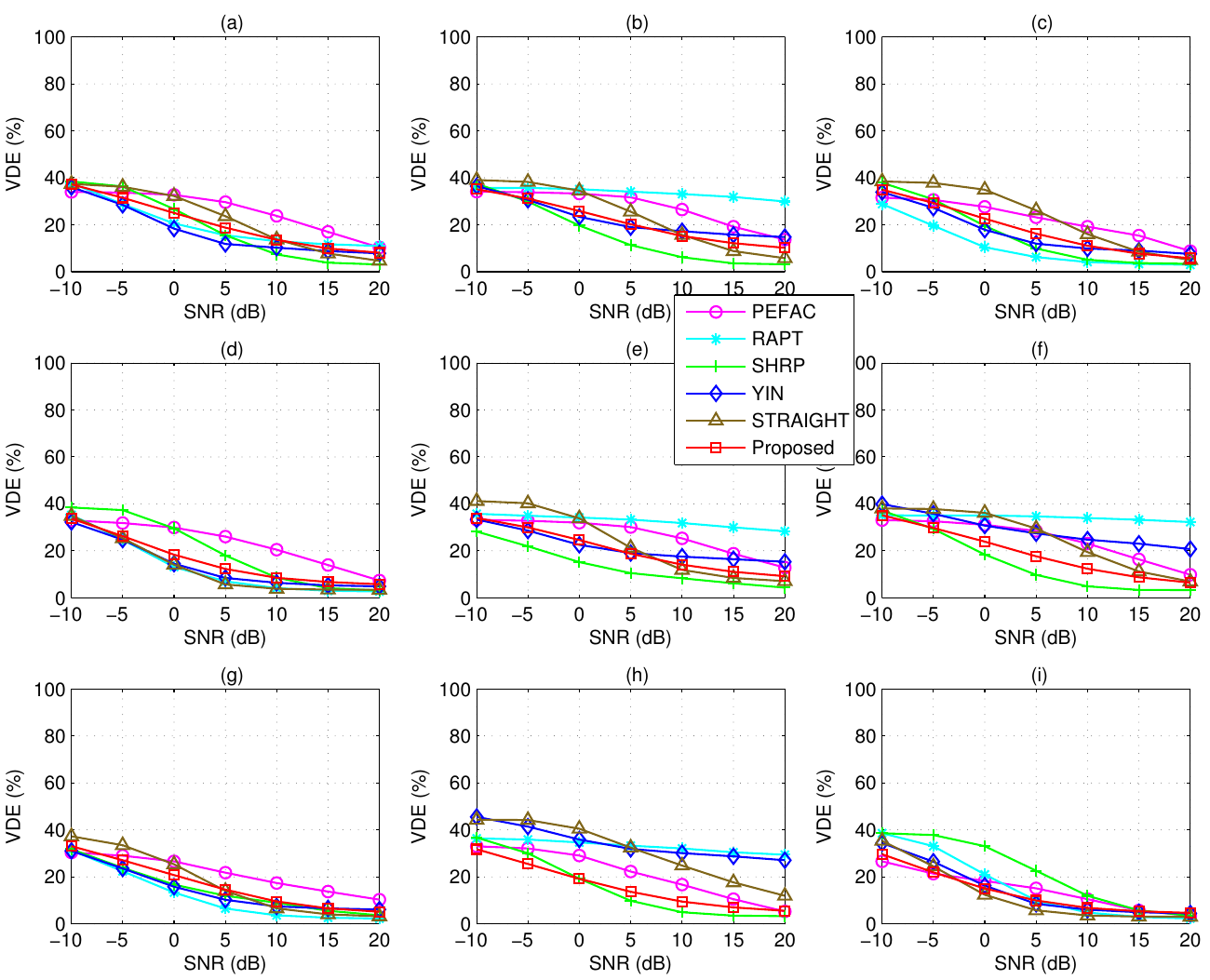}
\caption{VDE results using various methods for pitch estimation. Speech signals are from the CSTR database, while the additive noises are from the Aurora database and AWGN. Noise types are respectively (a) airport, (b) babble, (c) car, (d) exhibition, (e) restaurant, (f) street, (g) subway, (h) train, (i) AWGN.}
\label{fig:pitch_VDE}
\end{figure}

In Fig.~\ref{fig:pitch_VDE} we show the VDE results.
We can see that the VDE decreases as the SNR increases in general for all methods. 
The proposed method performs consistently in all the test cases. 
For high SNRs, the SHRP seems to perform the best.  
The RAPT seems to have difficulties with the ``speech-like'' noise types even at high SNRs (cf. Fig.~\ref{fig:singlespeakerPitch1} and Fig.~\ref{fig:singlespeakerPitch2}). 

\subsection{Reverberation}

\begin{figure}[!ht]
\centering
\includegraphics[width=0.49\textwidth]{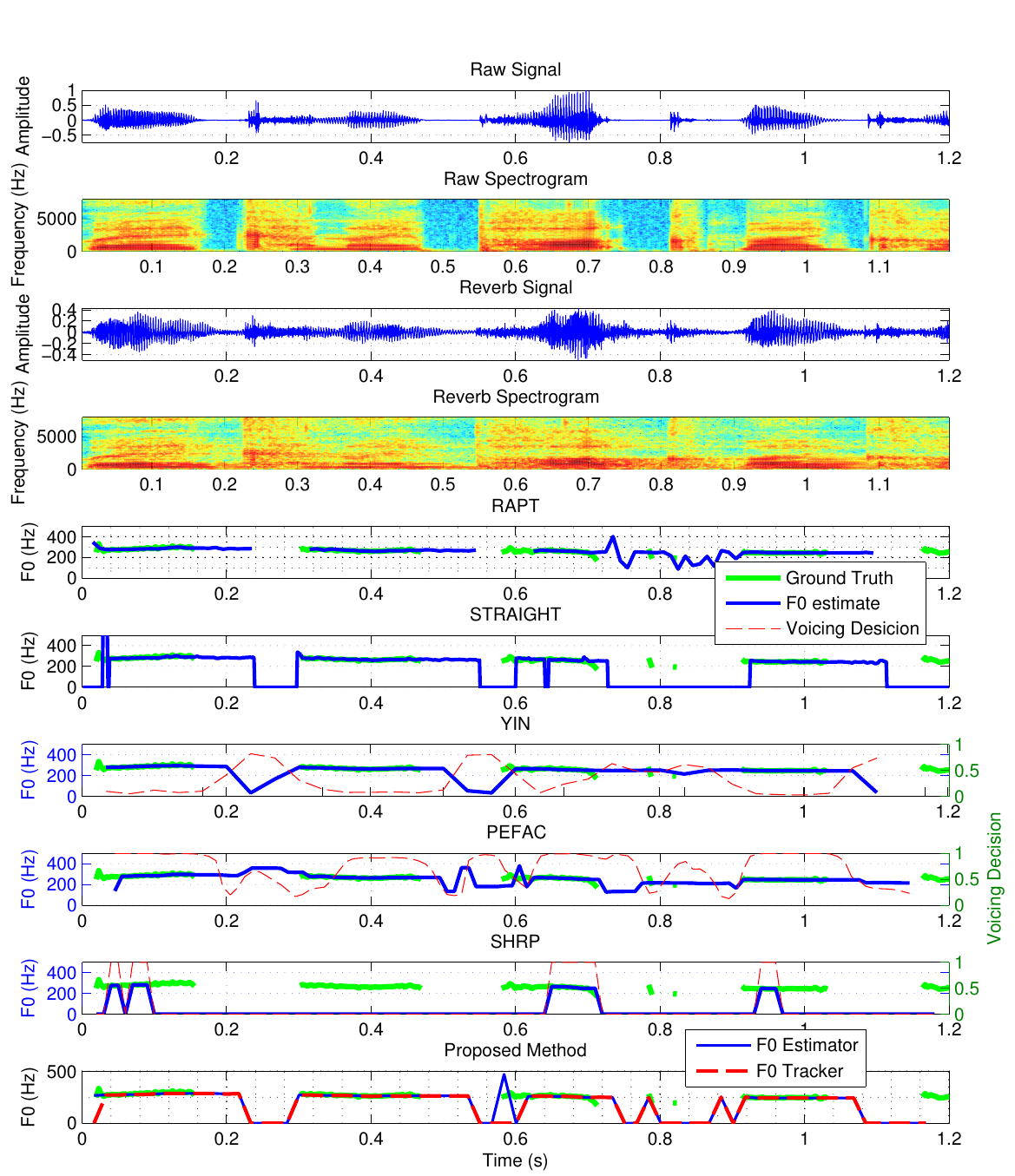}
\caption{Speech signals of a female speaker and its reverberant recording (\myedit{top four} panels)\myedit{, and pitch estimates from the RAPT, STRAIGHT, YIN, PEFAC, SHRP and the proposed methods respectively. No additive noise}. }
\label{fig:reverb1}
\end{figure}

Reverberation can also cause errors to most pitch estimation algorithms, because the reflections change the waveform and spectra of the sounds. 
In Fig.~\ref{fig:reverb1}, we show an example of real recordings of the speech signals from the CSTR database in an office room with reverberation time of $T_{60} \approx 0.65$s. {\color{black}A loudspeaker is used to play the original sound corpora, and the electret omnidirectional microphone, preamp and sound card are used for recording}. 
We can see that the reverberation creates long ``tails'' in the waveforms and spectra of the sound recording, especially during speech pauses, which is also obvious in the pitch estimation results using our proposed pitch estimator and tracker \myedit{(see the bottom panel)}, compared with the ground truth for the clean speech signal. However, the values of pitch estimates are close to the ground truth over time. 
%
%
Fig.~\myedit{\ref{fig:reverb1} also} shows the pitch estimation results from the RAPT, STRAIGHT, YIN, PEFAC and SHRP methods respectively, using the reverberant recording. 
We can see that all the methods (except the SHRP) produces ``tails'' due to the reverberation. 
The RAPT has some spurious estimates at around 0.8s. 
Estimates of the STRAIGHT overlap well with ground truth, except for the miss-detections at about 0.8s and 1.2s. 
The YIN and PEFAC, considering also their corresponding voicing decision measures, produce accurate pitch estimates. 
The SHRP however, has considerable miss-detections, although all its voiced estimates are accurate. 
%
%

\subsection{Multiple Speakers}

\begin{figure}[!h]
\centering
\includegraphics[width=0.49\textwidth]{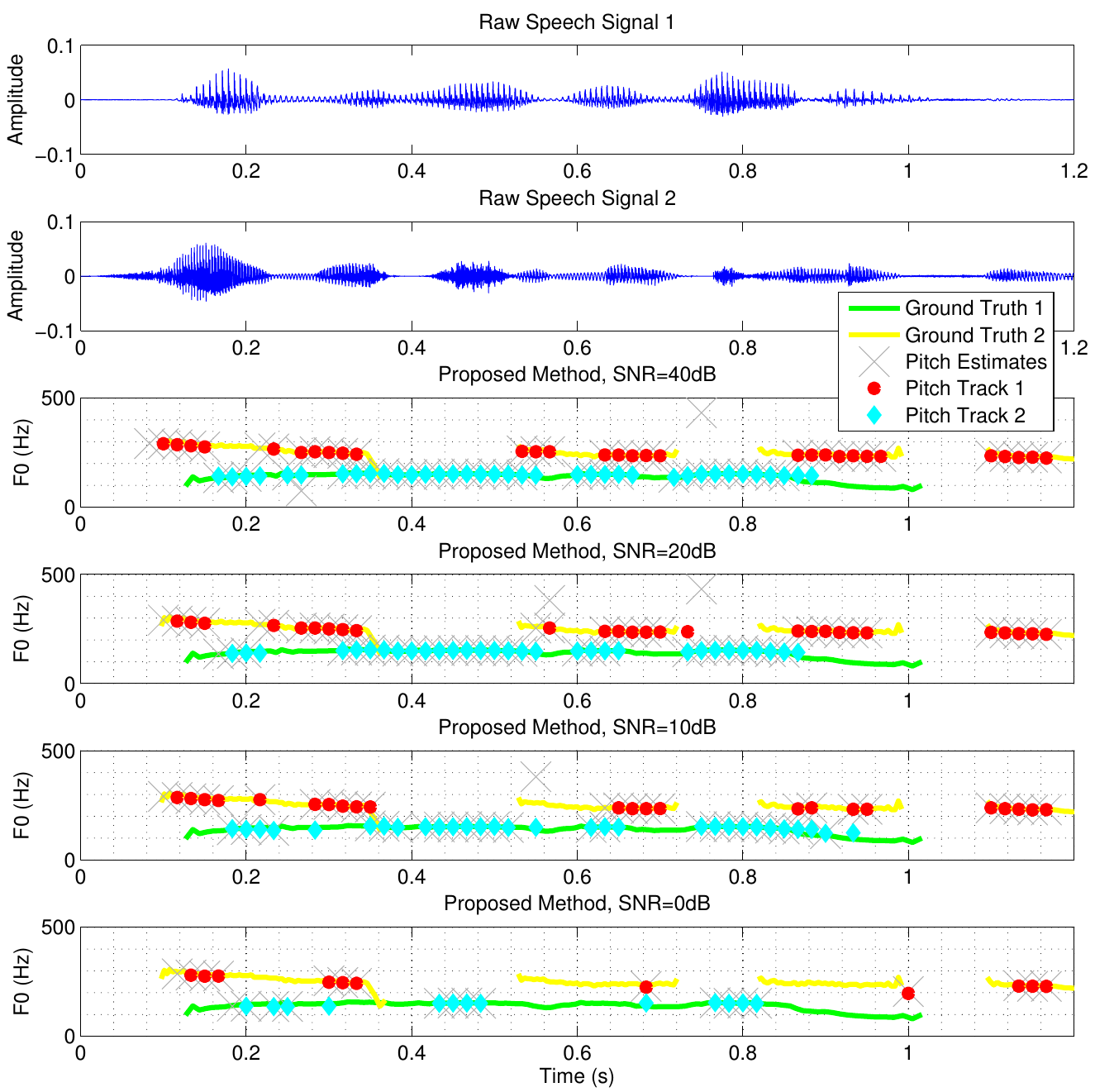}
\caption{Raw speech signals of a male speaker and a female speaker (top two panels), and pitch estimation {\color{black}and tracking} results of the mixture of the two concurrent speakers, with additive babble noise of SNR=40, 20, 10 and 0dB (bottom four panels). }
\label{fig:multipitch}
\end{figure}

We also test our proposed multi-pitch tracking method for concurrent speakers. 
\myedit{Since the GLMB filter is capable of tracking states of a time-varying number of objects, the proposed pitch tracker is applicable to the scenario with more than two concurrent speakers, provided they are at different pitch levels. In this paper, we focus on the scenario of two speakers as it is more common. 
}
Fig.~\ref{fig:multipitch} shows pitch estimation results from our proposed multi-pitch tracker for the case when a female speaker and a male speaker \myedit{(at different pitch levels)} talk concurrently. \myedit{The speech signals are chosen from the CSTR corpus and then normalized and superimposed.} Additive babble noise of various levels is included to test the reliability of the proposed method. 
We can see that at each SNR level, the proposed pitch tracker produces two separate pitch tracks and correctly assigns different labels (as shown in different colors) to the pitch estimates of the two respective speakers. 
There are unvoiced periods in the speech signals, but we can see that assigning the correct labels to the pitch estimates links the segments and forms a pitch track for each corresponding speaker. 
Moreover, the spurious pitch estimates from our proposed pitch estimator are filtered by the proposed pitch tracker since they do not form temporal continuity with their neighbouring pitch estimates.
When the noise is weak, the majority of the pitches are detected and most of the pitch estimates are accurate compared to the ground truth. 
As the noise gets stronger, there are more miss-detections. There are also brief miss-detections due to competing sounds, especially when the noise is strong. However, most of the pitch estimates from the proposed pitch tracker are still close to ground truth. 
%
%

%
\begin{figure}[h]
\centering
\includegraphics[width=0.49\textwidth]{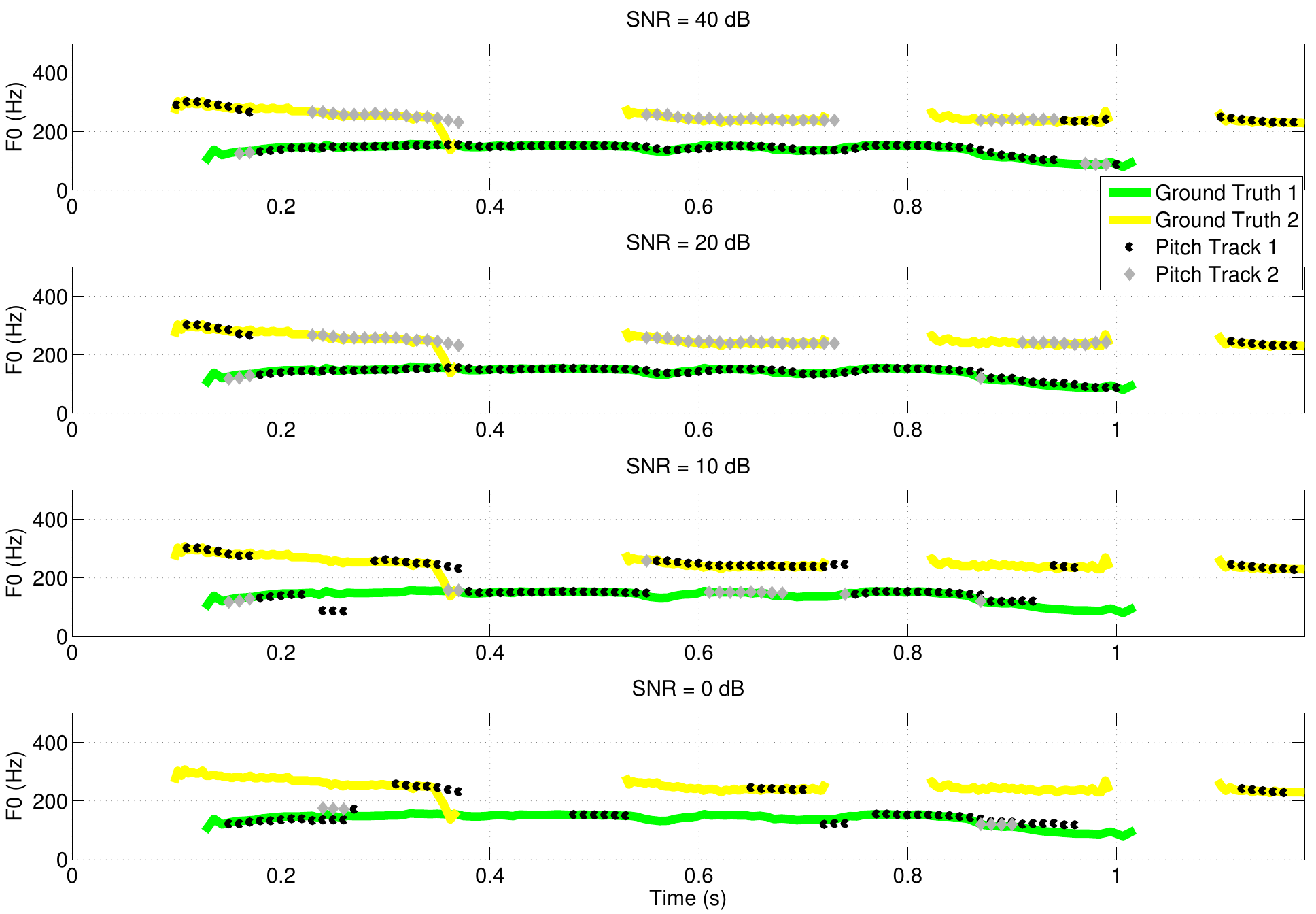}
\caption{Multi-pitch tracking results using Wu's method for the two concurrent speakers, with additive babble noise of SNR=40, 20, 10 and 0dB, respectively. }
\label{fig:WUmultipitch}
\end{figure}

For comparison, Fig.~\ref{fig:WUmultipitch} shows the multi-pitch tracking results of the same noisy speech mixtures at various SNRs, using Wu's method \cite{wu2003multipitch}. The pitch ground truth is plotted in each panel as reference. Wu's method produces two tracks of pitch estimates, i.e. Pitch Track 1 and Pitch Track 2, which are indicated with different colors. 
Here the available released C-code for Wu's method \cite{multipitchTrackingWu} is used as is, where the time step is $10$ms and the time frame length is $16$ms. Note that the results may not be the best possible from Wu's method, as ideally its parameters could be trained with the new database. We can see that at SNR=40dB, most of the pitches are extracted, but there are errors of pitch identities before $0.2$s and after $0.9$s. As the babble noise gets stronger, e.g. SNR=10dB and 0dB, there are more miss-detections and pitch identity errors, and almost all the identity of estimates of Pitch Track 2 are mistaken. 
We can also see in this case that the pitch identity errors of Wu's method do not always happen at the same time over different SNRs.

\begin{table}[!h] 
\centering
\caption{GPE of Wu's and the proposed multi-pitch tracking.}\label{gpeTab}
\begin{tabular}{ @{\extracolsep{5pt}} c c c c c @{} }
\hline \hline
& \multicolumn{2}{c} {Wu's } & 
	\multicolumn{2}{c} {Proposed } \\\cline{2-3}   \cline{4-5}
{SNR (dB)} & Speaker 1 & Speaker 2 & Speaker 1 & Speaker 2 \\
\hline
 40 & 0.2273 & 0.1333 & 0.0667 & 0.1081 \\  
 20 & 0.1705 & 0.1333 & 0.0385 & 0.0968 \\ 
 10 & 0.5625 & 0.9286 & 0 & 0.2333\\
 0  & 0.6364 & 1  & 0.1 & 0.3077\\
\hline \hline 
\end{tabular}
\end{table}
Taking into account errors on the pitch labeling (identities) as well as the pitch accuracy, Table~\ref{gpeTab} gives the GPE results for the multi-pitch tracking using the proposed method and the Wu's method.
We can see that the proposed method has less GPE compared with Wu's method over the range of SNRs, although in general the errors also tend to increase as the noise gets stronger. 
\myedit{Table~\ref{vdeTab} shows the corresponding VDE results, which indicates that the proposed method has more voicing decision errors (mostly miss-detections) than Wu's method for the multi-pitch scenario (cf. Fig.~\ref{fig:multipitch} and Fig.~\ref{fig:WUmultipitch}).}

\myedit{
\begin{table}[!h] 
\centering
\caption{VDE of Wu's and the proposed multi-pitch tracking.}\label{vdeTab}
\begin{tabular}{ @{\extracolsep{5pt}} c c c c c @{} }
\hline \hline
Methods \vline & SNR = 40 (dB) & 20 & 10 & 0 \\
\hline
 Wu's & 0.0085 & 0.0254 & 0.0847 & 0.3305 \\  
 Proposed & 0.0429 & 0.1143 & 0.2000 & 0.5143  \\ 
\hline \hline 
\end{tabular}
\end{table}
Accurately filtering raw pitch estimates and correctly associating estimates with respective speakers are crucial features of the multi-pitch tracker. 
Thus we also measure the speaker identity errors of the pitch trackers as defined in (\ref{eq:SIE}), i.e. for each speaker, the ratio between the number of pitch estimates that actually belong to another speaker (see e.g. the pitch estimates before 0.15s at SNR$=40$dB in Fig.~\ref{fig:WUmultipitch}), and its total number of pitch estimates. 
%
\begin{table}[!h] 
\centering
\caption{SIE of Wu's and the proposed multi-pitch tracking.}\label{labTab}
\begin{tabular}{ @{\extracolsep{5pt}} c c c c c @{} }
\hline \hline
& \multicolumn{2}{c} {Wu's } & 
	\multicolumn{2}{c} {Proposed } \\\cline{2-3}   \cline{4-5}
{SNR (dB)} & Speaker 1 & Speaker 2 & Speaker 1 & Speaker 2 \\
\hline
 40 & 0.1136 & 0.1111 & 0 & 0 \\  
 20 & 0.0568 & 0.0889 & 0 & 0 \\ 
 10 & 0.4500 & 0.9286 & 0 & 0\\
 0  & 0.2364 & 1  & 0 & 0\\
\hline \hline 
\end{tabular}
\end{table}
From Fig.~\ref{fig:multipitch} as expected, since the two speakers are at different pitch levels, there is zero $\mathrm{SIE}$ from the proposed method. 
However, from Fig.~\ref{fig:WUmultipitch}, there are considerable identity errors from Wu's method especially at low SNRs. The SIE results are provided in Table~\ref{labTab}. Overall, although having higher VDEs, the proposed method provides considerably better GPEs and SIEs in the studied scenario. 
}

\section{Conclusion}
\label{sec:conclusion}

In this paper we propose a new pitch estimator inspired by CASA approaches and a novel pitch tracker that does not require training. 
The pitch estimator uses an auditory filterbank  to decompose the speech mixture. The number of subbands and center frequencies of the filterbank are calculated according to our proposed frequency coverage metric for consistent and full frequency coverage without redundancy. For reliable and distinct pitch estimates, it encodes subband signals before the autocorrelation operation. 
To further suppress spurious errors and connect pitch tracks of respective speakers, the pitch tracker is proposed based on the GLMB framework, assuming temporal continuity of pitch. We propose a novel pitch transition model based on the Ornstein Uhlenbeck process, and use the measurement driven birth model for adaptive tracking.
We also provide some necessary adaptations of GLMB, including the single pitch tracking as well as the labeling in presence of unvoiced periods or long pauses during speech. 
Not only for single pitch tracking, this training-free pitch tracker is also applicable for multi-pitch tracking as long as pitches of concurrent speakers are on different levels.

The proposed pitch estimator and tracker produce not only reliable pitch estimates but also pitch labels (speaker identities) for respective speakers. 
Evaluations using the CSTR, Keele and AURORA databases as well as real recordings in a reverberant room of $T_{60} \approx 0.65s$ have demonstrated the reliability of the proposed methods against various additive noises and also reverberation. Numerical comparisons with other baseline methods also validate the benefits of the proposed methods.

\section*{Acknowledgment}
The author would like to acknowledge the contributions of the Australian Postgraduate Award and Australian Government Research Training Program Scholarship in supporting this research.
The author is grateful to Dr. DAn Ellis for the sound database that he generously shared and the intriguing discussions on the binary file conversion, 
to Dr. T.C. Toh for his helpful mathematical insights and review comments, 
and to anonymous reviewers for constructive comments that help improve the quality of the paper. 

%

%

\appendices

\section{Expression of the Subband Signal}
\label{appen:dirsubsig}

From (\ref{eq:speechModel}), the speech harmonic component is
\begin{equation}\label{eq:speechModelenv} 
\begin{aligned}
& s^{(\hbar)}_q(t) 
= A^{(\hbar)}_{q}(t) \cdot \cos \big( {\hbar} \cdot \omega_q \cdot t + \phi^{({\hbar})}_{q}(t) \big) \\
= & \frac{1}{2} {A}^{(\hbar)}_{q}(t) \cdot [ e^{\mathrm{i} [{\hbar} \cdot \omega_q \cdot t + \phi^{({\hbar})}_{q}(t)]} + e^{- \mathrm{i} [{\hbar} \cdot \omega_q \cdot t - \phi^{({\hbar})}_{q}(t)]} ]   
\end{aligned}
\end{equation}
where $\omega_q \triangleq 2\pi f_q$.%

Using linear-phase filters, e.g. the gammatone filter, from (\ref{eq:gammatoneenv}) we have
\begin{equation} \label{eq:gammatoneenvexp}
\begin{aligned}
g^{(b)}(t) 
& = \tilde{g}^{(b)}(t) \cdot \cos(2\pi f_C^{(b)} t ) \\
& = \frac{1}{2} \cdot \tilde{g}^{(b)}(t) \cdot ( e^{\mathrm{i} 2\pi f_C^{(b)} t} +  e^{- \mathrm{i} 2\pi f_C^{(b)} t} ) . 
\end{aligned}
\end{equation}

From (\ref{eq:speechModel}), (\ref{eq:gammatoneenv}) and (\ref{eq:RIR}), when $ {\hbar} \cdot \omega_q \approx 2\pi f_C^{(b)} $, the subband signal is given as follows: 
\begin{equation} \label{eq:pitchcosinederive}
\begin{aligned}
\allowbreak
 & {x}^{(b)}(t)
\approx  [ {s}^{(\hbar)}_q(t-t_{d_{q}})\cdot \mathrm{h}_{qi}(t_{d_{q}}) ]  \ast g^{(b)}(t)  \\
& = \mathrm{h}_{q}(t_{d_{q}}) \cdot  [ \frac{1}{2} \cdot \tilde{g}^{(b)}(t) \cdot ( e^{\mathrm{i} 2\pi f_C^{(b)} t} +  e^{- \mathrm{i} 2\pi f_C^{(b)} t} ) ] \\
&~~~  \ast  [  \frac{1}{2} {A}^{(\hbar)}_{q}(t-t_{d_{q}}) \\ &~ \cdot [ e^{\mathrm{i} [{\hbar} \cdot \omega_q \cdot (t-t_{d_{q}}) + \phi^{({\hbar})}_{q}(t-t_{d_{q}})]} + e^{- \mathrm{i} [{\hbar} \cdot \omega_q \cdot (t-t_{d_{q}}) - \phi^{({\hbar})}_{q}(t-t_{d_{q}})]} ]  \\
& \approx \frac{1}{4} \mathrm{h}_{q}(t_{d_{q}}) \cdot    
[ {A}^{(\hbar)}_{q}(t-t_{d_{q}}) \ast \tilde{g}^{(b)}(t) ] \\ & ~ \cdot [ e^{\mathrm{i} [{\hbar} \cdot \omega_q \cdot (t-t_{d_{q}}) + \phi^{({\hbar})}_{q}(t-t_{d_{q}})] } +   e^{- \mathrm{i} [{\hbar} \cdot \omega_q \cdot (t-t_{d_{q}}) - \phi^{({\hbar})}_{q}(t-t_{d_{q}}) ]} ]  \text{$^\dagger$ } \\
& =  \frac{1}{2} \mathrm{h}_{q}(t_{d_{q}}) \cdot     [ A^{(\hbar)}_{q}(t-t_{d_{q}}) \ast \tilde{g}^{(b)}(t)  ] \cdot \\
&~~~~~~ \cos ( {\hbar}  \omega_q  (t-t_{d_{q}})  + {\phi}_{q}^{({\hbar})}(t-t_{d_{q}})  )  \\
& =   \tilde{S}_{q}^{(b)}(t)  \cdot \cos ( \tilde{\phi}^{({b})}_{q}(t)  ) ,~ t \geq t_{{q}} ,
\end{aligned}
\footnotetext{$\dagger$ Considering frequency domain meanings of convolution and complex exponentials for the Fourier transform and inverse Fourier transform.}
\end{equation}
where
\begin{equation} \label{eq:pitchcosineenv}
\tilde{S}_{q}^{(b)}(t) \triangleq 
 \frac{1}{2} \cdot \mathrm{h}_{q}(t_{d_{q}})  \cdot 
  A^{(\hbar)}_{q}(t-t_{d_{q}}) \ast \tilde{g}^{(b)}(t)  , 
\end{equation}
and
\begin{equation} \label{eq:pitchphaseCosine}
\tilde{\phi}^{({b})}_{q}(t) 
=  2 \pi {\hbar}  f_q  (t-t_{d_{q}})  + \phi^{({\hbar})}_{q}(t-t_{d_{q}}) . 
\end{equation}
\hspace*{\fill}$\blacksquare$

{\color{black}
\section{Derivation of the ERBS Expression}
\label{sec:ERBS}
\begin{proof}
From (\ref{eq:ERBS}),
\begin{equation}
\begin{aligned} \label{eq:ERBSproof}
& {\Upsilon}(f) \triangleq \int \frac{1}{{\upsilon}(f)} df = \int \frac{1}{D+E\cdot f}df \\
= & \frac{1}{E}\cdot \ln (D+E\cdot f) + \mathrm{Constant} \\
= & \frac{1}{E\cdot \lg e} \lg (D+E\cdot f) + \mathrm{Constant} \\
= & \frac{1}{E\cdot \lg e} [\lg (1+\frac{E}{D}\cdot f) + \lg D] + \mathrm{Constant} ,
\end{aligned}
\end{equation}
\myedit{which} using the boundary condition that ${\Upsilon}(0) = 0$, leads to ${\Upsilon}(f)= \frac{1}{E\cdot \lg e} [\lg (1+\frac{E}{D}\cdot f)]$, hence (\ref{eq:ERBS}).
\end{proof}
}

\section{Frequency Coverage and Number of Subbands}
\label{sec:appendix2}
\begin{proof}
From (\ref{eq:etaCb}), (\ref{eq:Kvartheta}) and (\ref{eq:upsilon0}), the frequency coverage is 
\allowbreak
\begin{equation} \label{eq:etaCpsi}
\begin{aligned}
& \eta_{C}^{(b)} \triangleq \frac{1}{2} \cdot \frac{ f_B^{(b+1)} + f_B^{(b)} }{ f_C^{(b+1)} - f_C^{(b)} } \\
& = K_\vartheta \cdot \frac{ D + \frac{E}{2} \cdot ( f_C^{(b+1)} + f_C^{(b)} ) }{ f_C^{(b+1)} - f_C^{(b)} } \\
& = \frac{E \!\! \cdot \!\! K_\vartheta }{2} \cdot \Big[ \Big( (1+D' f_{min})^{(N_b-b-1)} \!\! \cdot \!\! (1+D' f_{max})^{(b)} \Big) ^{\frac{1}{N_b-1}} \!\! + \\
& ~~~~~~~ \Big( (1+D' f_{min})^{(N_b-b)} \cdot (1+D' f_{max})^{(b-1)} \Big) ^{\frac{1}{N_b-1}}   \Big] \Big/ \\ 
&~~~~~~~~ \Big[ \Big( (1+D' f_{min})^{(N_b-b-1)} \cdot (1+D' f_{max})^{(b)} \Big) ^{\frac{1}{N_b-1}} - \\
&~~~~~~~~~ \Big( (1+D' f_{min})^{(N_b-b)} \cdot (1+D' f_{max})^{(b-1)} \Big) ^{\frac{1}{N_b-1}}   \Big] \\
& = \frac{E \cdot K_\vartheta }{2} \cdot \Big[ (1+D' f_{max})^{\frac{1}{N_b-1}}   +  (1+D' f_{min})^{\frac{1}{N_b-1}}   \Big] \Big/ \\ 
&~~~~~~ \Big[ (1+D' f_{max})^{\frac{1}{N_b-1}}  - (1+D' f_{min})^{\frac{1}{N_b-1}}   \Big] \\
& = \frac{E \cdot K_\vartheta }{2} \cdot \frac{(\frac{D + E\cdot f_{max}}{D + E \cdot f_{min}})^{\frac{1}{N_b-1}}+1}{(\frac{D + E \cdot f_{max}}{D + E \cdot f_{min}})^{\frac{1}{N_b-1}}-1} ,
\end{aligned}
\end{equation}
where $K_\vartheta$ for the gammatone filter is a constant for a given filter order $\vartheta$\cite{holdsworth1988implementing}, 
\begin{equation} \label{eq:Ktheta}
K_\vartheta = 2 \sqrt{2^{1/\vartheta}-1} \cdot 
\Big[ \frac{\pi (2\vartheta-2)!2^{-(2\vartheta-2)}}{(\vartheta-1)!^2} \Big]^{-1} .
\end{equation}
Thus the number of subbands $N_b$ as given in (\ref{eq:Nb}) can be directly obtained from (\ref{eq:etaCpsi}).
\end{proof}

\ifCLASSOPTIONcaptionsoff
  \newpage
\fi

\bibliographystyle{IEEEtran}

\bibliography{Bibliography}

@book{CASAwang,
  title={Computational auditory scene analysis: Principles, algorithms, and applications},
  author={Wang, DeLiang and Brown, Guy J},
  year={2006},
  publisher={Wiley-IEEE Press}
}

@book{deller1993discrete,
  title={Discrete time processing of speech signals},
  author={Deller Jr, John R and Proakis, John G and Hansen, John H},
  year={1993},
  publisher={Prentice Hall PTR}
}

@book{gardiner1985handbook,
  title={Handbook of stochastic methods},
  author={Gardiner, Crispin W and others},
  volume={3},
  year={1985},
  publisher={Springer Berlin}
}

@article{glasberg1990derivation,
  title={Derivation of auditory filter shapes from notched-noise data},
  author={Glasberg, Brian R and Moore, Brian CJ},
  journal={Hearing research},
  volume={47},
  number={1},
  pages={103--138},
  year={1990},
  publisher={Elsevier}
}

@article{meddis1986simulation,
  title={Simulation of mechanical to neural transduction in the auditory receptor},
  author={Meddis, Ray},
  journal={The Journal of the Acoustical Society of America},
  volume={79},
  number={3},
  pages={702--711},
  year={1986},
  publisher={Acoustical Society of America}
}

@inproceedings{patterson1987efficient,
  title={An efficient auditory filterbank based on the gammatone function},
  author={Patterson, RD and Nimmo-Smith, Ian and Holdsworth, John and Rice, Peter},
  booktitle={a meeting of the IOC Speech Group on Auditory Modelling at RSRE},
  volume={2},
  number={7},
  year={1987}
}

@article{yilmaz2004blind,
  title={Blind separation of speech mixtures via time-frequency masking},
  author={Yilmaz, Ozgur and Rickard, Scott},
  journal={IEEE Transactions on Signal Processing},
  volume={52},
  number={7},
  pages={1830--1847},
  year={2004},
  publisher={IEEE}
}

@book{mahler2007statistical,
  title={Statistical multisource-multitarget information fusion},
  author={Mahler, Ronald PS},
  year={2007},
  publisher={Artech House, Inc.}
}

@article{reuter2014labeled,
  title={The labeled multi-Bernoulli filter},
  author={Reuter, Stephan and Vo, Ba-Tuong and Vo, Ba-Ngu and Dietmayer, Klaus},
  journal={IEEE Transactions on Signal Processing},
  volume={62},
  number={12},
  pages={3246--3260},
  year={2014},
  publisher={IEEE}
}

@article{vo2014labeled,
  title={Labeled random finite sets and the Bayes multi-target tracking filter},
  author={Vo, Ba-Ngu and Vo, Ba-Tuong and Phung, Dinh},
  journal={IEEE Transactions on Signal Processing},
  volume={62},
  number={24},
  pages={6554--6567},
  year={2014},
  publisher={IEEE}
}

@article{vo2013labeled,
  title={Labeled random finite sets and multi-object conjugate priors},
  author={Vo, Ba-Tuong and Vo, Ba-Ngu},
  journal={IEEE Transactions on Signal Processing},
  volume={61},
  number={13},
  pages={3460--3475},
  year={2013},
  publisher={IEEE}
}

@article{varga1993assessment,
  title={Assessment for automatic speech recognition: II. NOISEX-92: A database and an experiment to study the effect of additive noise on speech recognition systems},
  author={Varga, Andrew and Steeneken, Herman JM},
  journal={Speech communication},
  volume={12},
  number={3},
  pages={247--251},
  year={1993},
  publisher={Elsevier}
}

@inproceedings{auroraNoise,
	author = {\text{Available} at https://www.ee.columbia.edu/$\sim$dpwe/sounds/noise/}
	}

@inproceedings{fdaSpeech,
	author = {\text{Available} at http://www.cstr.ed.ac.uk/research/projects/fda/}
	}

@inproceedings{plante1995pitch,
  title={A pitch extraction reference database},
  author={Plante, Fabrice and Meyer, Georg F and Ainsworth, William A},
  booktitle={Fourth European Conference on Speech Communication and Technology},
  year={1995}
}

@article{smith1999bark,
  title={Bark and ERB bilinear transforms},
  author={Smith III, Julius O and Abel, Jonathan S},
  journal={IEEE Transactions on Speech and Audio Processing},
  volume={7},
  number={6},
  pages={697--708},
  year={1999},
  publisher={IEEE}
}

@inproceedings{lyon1983computational,
  title={A computational model of binaural localization and separation},
  author={Lyon, Richard},
  booktitle={Acoustics, Speech, and Signal Processing, IEEE International Conference on ICASSP'83.},
  volume={8},
  pages={1148--1151},
  year={1983},
  organization={IEEE}
}

@article{de2002yin,
  title={YIN, a fundamental frequency estimator for speech and music},
  author={De Cheveign{\'e}, Alain and Kawahara, Hideki},
  journal={The Journal of the Acoustical Society of America},
  volume={111},
  number={4},
  pages={1917--1930},
  year={2002},
  publisher={Acoustical Society of America}
}

@article{tolonen2000computationally,
  title={A computationally efficient multipitch analysis model},
  author={Tolonen, Tero and Karjalainen, Matti},
  journal={IEEE transactions on speech and audio processing},
  volume={8},
  number={6},
  pages={708--716},
  year={2000},
  publisher={IEEE}
}

@article{wu2003multipitch,
  title={A multipitch tracking algorithm for noisy speech},
  author={Wu, Mingyang and Wang, DeLiang and Brown, Guy J},
  journal={IEEE Transactions on Speech and Audio Processing},
  volume={11},
  number={3},
  pages={229--241},
  year={2003},
  publisher={IEEE}
}

@inproceedings{multipitchTrackingWu,
	author = {\text{Available at http://web.cse.ohio-state.edu/pnl/shareware/wu-tsap03/}}
	}

@inproceedings{noll1970pitch,
  title={Pitch determination of human speech by the harmonic product spectrum, the harmonic surn spectrum, and a maximum likelihood estimate},
  author={Noll, A Michael},
  booktitle={Symposium on Computer Processing in Communication, ed.},
  volume={19},
  pages={779--797},
  year={1970},
  organization={University of Broodlyn Press, New York}
}

@article{ross1974average,
  title={Average magnitude difference function pitch extractor},
  author={Ross, M and Shaffer, H and Cohen, Andrew and Freudberg, Richard and Manley, H},
  journal={IEEE Transactions on Acoustics, Speech, and Signal Processing},
  volume={22},
  number={5},
  pages={353--362},
  year={1974},
  publisher={IEEE}
}

@article{hermes1988measurement,
  title={Measurement of pitch by subharmonic summation},
  author={Hermes, Dik J},
  journal={The journal of the acoustical society of America},
  volume={83},
  number={1},
  pages={257--264},
  year={1988},
  publisher={ASA}
}

@article{maragos1993energy,
  title={Energy separation in signal modulations with application to speech analysis},
  author={Maragos, Petros and Kaiser, James F and Quatieri, Thomas F},
  journal={IEEE transactions on signal processing},
  volume={41},
  number={10},
  pages={3024--3051},
  year={1993},
  publisher={IEEE}
}

@article{meddis1997unitary,
  title={A unitary model of pitch perception},
  author={Meddis, Ray and O’Mard, Lowel},
  journal={The Journal of the Acoustical Society of America},
  volume={102},
  number={3},
  pages={1811--1820},
  year={1997},
  publisher={ASA}
}

@article{christensen2007joint,
  title={Joint high-resolution fundamental frequency and order estimation},
  author={Christensen, Mads Grsbll and Jakobsson, Andreas and Jensen, Sren Holdt},
  journal={IEEE Transactions on Audio, Speech, and Language Processing},
  volume={15},
  number={5},
  pages={1635--1644},
  year={2007},
  publisher={IEEE}
}

@inproceedings{christensen2008robust,
  title={Robust subspace-based fundamental frequency estimation},
  author={Christensen, Mads G and Vera-Candeas, Pedro and Somasundaram, Samuel D and Jakobsson, Andreas},
  booktitle={IEEE International Conference on Acoustics, Speech and Signal Processing, 2008. ICASSP 2008. },
  pages={101--104},
  year={2008},
  organization={IEEE}
}

@article{zhang2010robust,
  title={A robust and computationally efficient subspace-based fundamental frequency estimator},
  author={Zhang, Johan Xi and Christensen, Mads Gr{\ae}sb{\o}ll and Jensen, S{\o}ren Holdt and Moonen, Marc},
  journal={IEEE transactions on audio, speech, and language processing},
  volume={18},
  number={3},
  pages={487--497},
  year={2010},
  publisher={IEEE}
}

@article{tabrikian2004maximum,
  title={Maximum a-posteriori probability pitch tracking in noisy environments using harmonic model},
  author={Tabrikian, Joseph and Dubnov, Shlomo and Dickalov, Yulya},
  journal={IEEE Transactions on Speech and Audio Processing},
  volume={12},
  number={1},
  pages={76--87},
  year={2004},
  publisher={IEEE}
}

@article{wohlmayr2011probabilistic,
  title={A probabilistic interaction model for multipitch tracking with factorial hidden Markov models},
  author={Wohlmayr, Michael and Stark, Michael and Pernkopf, Franz},
  journal={IEEE Transactions on Audio, Speech, and Language Processing},
  volume={19},
  number={4},
  pages={799--810},
  year={2011},
  publisher={IEEE}
}

@article{holdsworth1988implementing,
  title={Implementing a gammatone filter bank},
  author={Holdsworth, John and Nimmo-Smith, Ian and Patterson, Roy and Rice, Peter},
  journal={Annex C of the SVOS Final Report: Part A: The Auditory Filterbank},
  volume={1},
  pages={1--5},
  year={1988}
}

@article{rouat1997pitch,
  title={A pitch determination and voiced/unvoiced decision algorithm for noisy speech},
  author={Rouat, Jean and Liu, Yong Chun and Morissette, Daniel},
  journal={Speech Communication},
  volume={21},
  number={3},
  pages={191--207},
  year={1997},
  publisher={Elsevier}
}

@inproceedings{sun2002pitch,
  title={Pitch determination and voice quality analysis using subharmonic-to-harmonic ratio},
  author={Sun, Xuejing},
  booktitle={2002 IEEE International Conference on Acoustics, Speech, and Signal Processing (ICASSP) },
  volume={1},
  pages={I--333},
  year={2002},
  organization={IEEE}
}

@article{gonzalez2014pefac,
  title={PEFAC-a pitch estimation algorithm robust to high levels of noise},
  author={Gonzalez, Sira and Brookes, Mike},
  journal={IEEE/ACM Transactions on Audio, Speech, and Language Processing},
  volume={22},
  number={2},
  pages={518--530},
  year={2014},
  publisher={IEEE}
}

@inproceedings{nolan2003intonational,
  title={Intonational equivalence: an experimental evaluation of pitch scales},
  author={Nolan, Francis},
  booktitle={Proceedings of the 15th International Congress of Phonetic Sciences, Barcelona},
  volume={39},
  year={2003}
}

@inproceedings{lee2012noise,
  title={Noise robust pitch tracking by subband autocorrelation classification},
  author={Lee, Byung Suk and Daniel P.W. Ellis},
  booktitle={Proc. INTERSPEECH, Portland, OR, USA},
  year={Sep, 2012},
}

@article{chu2012safe,
  title={SAFE: A statistical approach to F0 estimation under clean and noisy conditions},
  author={Chu, Wei and Alwan, Abeer},
  journal={IEEE Transactions on Audio, Speech, and Language Processing},
  volume={20},
  number={3},
  pages={933--944},
  year={2012},
  publisher={IEEE}
}

@inproceedings{lin2016measurement,
  title={Measurement driven birth model for the generalized labeled multi-Bernoulli filter},
  author={Lin, Shoufeng and Vo, Ba Tuong and Nordholm, Sven E},
  booktitle={2016 International Conference on Control, Automation and Information Sciences (ICCAIS)},
  pages={94--99},
  year={2016},
  publisher={IEEE}
}

@article{wang2017robust,
  title={Robust harmonic features for classification-based pitch estimation},
  author={Wang, Dongmei and Yu, Chengzhu and Hansen, John HL},
  journal={IEEE/ACM Transactions on Audio, Speech, and Language Processing},
  volume={25},
  number={5},
  pages={952--964},
  year={2017},
  publisher={IEEE}
}

@article{bagshaw1993enhanced,
  title={Enhanced pitch tracking and the processing of f0 contours for computer aided intonation teaching.},
  author={Bagshaw, Paul C and Hiller, Steven M and Jack, Mervyn A},
  year={1993},
  journal={Proc. Eurospeech},
  pages={1003--1006},
  publisher={International Speech Communication Association}
}

@article{talkin1995robust,
  title={A robust algorithm for pitch tracking (RAPT)},
  author={Talkin, David},
  journal={Speech coding and synthesis},
  volume={495},
  pages={518},
  year={1995}
}

@article{lin2018new,
  title={A New Frequency Coverage Metric and A New Subband Encoding Model, With An Application In Pitch Estimation},
  author={Lin, Shoufeng},
  journal={Proc. Interspeech 2018},
  pages={2147--2151},
  year={2018}
}

@article{lin2018reverberation,
  title={Reverberation-Robust Localization of Speakers Using Distinct Speech Onsets and Multichannel Cross Correlations},
  author={Lin, Shoufeng},
  journal={IEEE/ACM Transactions on Audio, Speech and Language Processing (TASLP)},
  volume={26},
  number={11},
  pages={2098--2111},
  year={2018},
  publisher={IEEE Press}
}

@inproceedings{kawahara1999fixed,
  title={Fixed point analysis of frequency to instantaneous frequency mapping for accurate estimation of F0 and periodicity},
  author={Kawahara, Hideki and Katayose, Haruhiro and Cheveign{\'e}, Alain de and Patterson, Roy D},
  booktitle={Sixth European Conference on Speech Communication and Technology},
  year={1999}
}

@article{kawahara1999restructuring,
  title={Restructuring speech representations using a pitch-adaptive time--frequency smoothing and an instantaneous-frequency-based F0 extraction: Possible role of a repetitive structure in sounds},
  author={Kawahara, Hideki and Masuda-Katsuse, Ikuyo and De Cheveigne, Alain},
  journal={Speech communication},
  volume={27},
  number={3-4},
  pages={187--207},
  year={1999},
  publisher={Elsevier}
}

@article{garner2013simple,
  title={A simple continuous pitch estimation algorithm},
  author={Garner, Philip N and Cernak, Milos and Motlicek, Petr},
  journal={IEEE Signal Processing Letters},
  volume={20},
  number={1},
  pages={102--105},
  year={2013},
  publisher={IEEE}
}

@inproceedings{wang2014f0,
  title={F0 estimation in noisy speech based on long-term harmonic feature analysis combined with neural network classification},
  author={Wang, Dongmei and Loizou, Philipos C and Hansen, John HL},
  booktitle={Fifteenth Annual Conference of the International Speech Communication Association},
  year={2014}
}

@inproceedings{zhang2016rnn,
  title={RNN-BLSTM Based Multi-Pitch Estimation.},
  author={Zhang, Jianshu and Tang, Jian and Dai, Li-Rong},
  booktitle={INTERSPEECH},
  pages={1785--1789},
  year={2016}
}

@article{reddy2017robust,
  title={Robust pitch extraction method for the HMM-based speech synthesis system},
  author={Reddy, M Kiran and Rao, K Sreenivasa},
  journal={IEEE Signal Processing Letters},
  volume={24},
  number={8},
  pages={1133--1137},
  year={2017},
  publisher={IEEE}
}

@article{liu2017speaker,
  title={Speaker-dependent multipitch tracking using deep neural networks},
  author={Liu, Yuzhou and Wang, DeLiang},
  journal={The Journal of the Acoustical Society of America},
  volume={141},
  number={2},
  pages={710--721},
  year={2017},
  publisher={ASA}
}

\end{document}